\documentclass[twocolumn]{aastex7}

\usepackage{upgreek}

\newcommand{\angstrom}{\textup{\AA} }
\newcommand{\angstromns}{\textup{\AA}}
\newcommand{\mic}{$\upmu$m }
\newcommand{\micns}{$\upmu$m}

\newcommand{\nev}{\mbox{[Ne \textsc{v}] }}
\newcommand{\nevns}{\mbox{[Ne \textsc{v}]}}
\newcommand{\net}{\mbox{[Ne \textsc{iii}] }}
\newcommand{\ot}{\mbox{[O \textsc{iii}] }}

\newcommand{\of}{\mbox{[O \textsc{iv}] }}

\newcommand{\netns}{\mbox{[Ne \textsc{iii}]}}
\newcommand{\otns}{\mbox{[O \textsc{iii}]}}

\newcommand{\netw}{\mbox{[Ne \textsc{ii}] }}

\newcommand{\st}{\mbox{[S \textsc{iii}] }}

\newcommand{\sfo}{\mbox{[S \textsc{iv}] }}
\newcommand{\sfons}{\mbox{[S \textsc{iv}]}}
\newcommand{\sn}{\mbox{[S \textsc{ix}] }}
\newcommand{\snns}{\mbox{[S \textsc{ix}]}}
\newcommand{\sisev}{\mbox{[Si \textsc{vii}] }}

\newcommand{\mgeig}{\mbox{[Mg \textsc{viii}] }}
\newcommand{\mgeigns}{\mbox{[Mg \textsc{viii}]}}
\newcommand{\feeig}{\mbox{[Fe \textsc{viii}] }}
\newcommand{\feeigns}{\mbox{[Fe \textsc{viii}]}}
\newcommand{\mgsev}{\mbox{[Mg \textsc{vii}] }}
\newcommand{\mgsevns}{\mbox{[Mg \textsc{vii}]}}
\newcommand{\nesix}{\mbox{[Ne \textsc{vi}] }}
\newcommand{\nesixns}{\mbox{[Ne \textsc{vi}]}}

\newcommand{\osev}{\mbox{O \textsc{vii} (f) }}
\newcommand{\osevns}{\mbox{O \textsc{vii} (f)}}

\newcommand{\nitsix}{\mbox{N \textsc{vi} (f) }}
\newcommand{\nitsixns}{\mbox{N \textsc{vi} (f)}}

\newcommand{\csix}{\mbox{C \textsc{vi} Ly$\alpha$ }}
\newcommand{\csixns}{\mbox{C \textsc{vi} Ly$\alpha$}}

\usepackage{siunitx}
\usepackage{amsmath}
\usepackage{multirow}
\usepackage{enumitem}
\usepackage{xcolor}
\usepackage[T1]{fontenc}

\accepted{\today}
\submitjournal{ApJ}

\begin{document}

\title{Footprints in the Wind: Probing X-ray Outflows in NGC~7469 using Near-Infrared Emission Lines}

\author[orcid=0000-0002-5718-2402,sname='Feuillet']{Léa M. Feuillet}
\affiliation{Institute for Astrophysics and Computational Sciences, Department of Physics, The Catholic University of America, Washington, DC 20064, USA}
\email[show]{feuilletl@cua.edu}  

\author[orcid=0000-0003-4073-8977,sname='Kraemer']{Steve Kraemer}
\affiliation{Institute for Astrophysics and Computational Sciences, Department of Physics, The Catholic University of America, Washington, DC 20064, USA}
\email{kraemer@cua.edu}

\author[orcid=0000-0001-8112-3464, sname='Trindade Falcão']{Anna Trindade Falcão}
\affiliation{NASA Goddard Space Flight Center, Greenbelt, MD, 20771, USA}
\email{anna.l.trindadefalcao@nasa.gov}

\author[orcid=0000-0002-2629-4989, sname='Braito']{Valentina Braito}
\affiliation{Institute for Astrophysics and Computational Sciences, Department of Physics, The Catholic University of America, Washington, DC 20064, USA}
\affiliation{INAF, Osservatorio Astronomico di Brera, Via Bianchi 46 I-23807 Merate (LC), Italy}
\affiliation{Dipartimento di Fisica, Università di Trento, Via Sommarive 14, Trento 38123, Italy}
\email{braito@cua.edu}

\author[orcid=0000-0001-8485-0325,sname='Meléndez']{Marcio Meléndez}
\affiliation{Space Telescope Science Institute, 3700 San Martin Drive, Baltimore, MD 21218, USA}
\email{melendez@stsci.edu}

\author[orcid=0000-0003-2450-3246,sname='Schmitt']{Henrique R. Schmitt}
\affiliation{Naval Research Laboratory, Remote Sensing Division, 4555 Overlook Ave SW, Washington, DC 20375, USA}
\email{schmitt.henrique@gmail.com}

\author[orcid=0000-0002-0982-0561,sname='Reeves']{James N. Reeves}
\affiliation{Institute for Astrophysics and Computational Sciences, Department of Physics, The Catholic University of America, Washington, DC 20064, USA}
\email{reeves@cua.edu}

\author[orcid=0000-0001-9815-9092, sname='Middei']{Riccardo Middei}
\affiliation{INAF Osservatorio Astronomico di Roma, Via Frascati 33, 00078 Monte Porzio Catone, RM, Italy}
\affiliation{Space Science Data Center, Agenzia Spaziale Italiana, Via del Politecnico snc, 00133 Roma, Italy}
\email{riccardo.middei@inaf.it}

\author[orcid=0000-0002-3365-8875, sname='Fischer']{Travis C. Fischer}
\affiliation{AURA for ESA, Space Telescope Science Institute, 3700 San Martin Drive, Baltimore, MD 21218, USA}
\email{tfischer@stsci.edu}

\author[orcid=0000-0002-4917-7873, sname='Revalski']{Mitchell Revalski}
\affiliation{Space Telescope Science Institute, 3700 San Martin Drive, Baltimore, MD 21218, USA}
\email{mrevalski@stsci.edu}

\begin{abstract}

AGN winds play an important role in the co-evolution of supermassive black holes and their host galaxies, yet their driving mechanisms and impact on star formation remain subjects of active investigation. Critically, the lack of X-ray Integral Field Units currently limits our ability to acquire spatially resolved velocity information in the X-ray regime. However, instead, this can be achieved using the James Webb Space Telescope. As part of an ongoing investigation of the nuclear feedback processes in the nearby luminous AGN NGC~7469, we present an analysis of the kinematics \added{of the X-ray emitting outflows} using near-infrared footprint lines such as \mgeig $\lambda$3.03~\micns. These high-ionization emission lines are associated with the same gas analyzed in the X-ray, and thus can be used to probe the footprint of the X-ray wind’s velocity structure and ionization state. Thanks to the wide wavelength range available with JWST we also use nebular (e.g. \sfo $\lambda$10.51~\micns) and coronal (e.g. \nev $\lambda$14.32~\micns) emission lines to offer a comprehensive multi-phase view of the outflows. We present mass and kinetic energy outflow rates, and find that while the feedback processes in NGC~7469 are not efficient by theoretical benchmarks, the most massive and energetic component is the high ionization X-ray gas.
\end{abstract}



\section{Introduction} 

High-ionization lines with ionization potentials (IPs) greater than 100eV, also known as coronal lines, have been detected in the spectra of Active Galactic Nuclei (AGN) since the 1980s (\citealp{Penston1984, Ward1984}, see \citealp{Rodriguez2025} for a review). These lines have been shown to be reliable tracers of AGN activity \citep{RodriguezArdila2002, Kynoch2022}. More recently, \citet{Trindade2022} introduced the term``footprint lines" to describe a subset of high-ionization emission lines in the optical and infrared (IR) with IPs greater than that of [O \textsc{vii}] (138 eV) that can be used to trace soft X-ray outflows in AGN. Winds are widespread among AGN \citep{Mullaney2013, Woo2016, Torres2020}, and have the ability to affect the host galaxy through a process referred to as feedback. By studying the kinematics and spatial distribution of the footprint lines, we can gain insight into potential winds and outflows from the AGN and their impact on nearby star formation.

NGC~7469 is a Seyfert 1.5 galaxy \citep{Veron2006}, which contains both narrow and broad components to H$\beta$. It is a nearby AGN (70.1 Mpc away, or z = 0.016268; \citealp{Springbob2005}) which allows us to resolve both the nucleus and the surrounding star formation. In fact, in addition to star formation in the host, NGC~7469 contains a ring of intense star formation surrounding the AGN with a radius of approximately 500 pc or 1\farcs5 \citep{Genzel1995}.

NGC~7469 has been extensively studied in the X-ray regime since the first XMM-Newton observations in 2000. Research has primarily focused on two key areas: X-ray variability and photoionization modeling of the warm absorbers (WAs). The galaxy exhibits variability on timescales as short as days \citep{nandra_origin_2000, partington_testing_2025}. Although the X-ray continuum varies, no significant changes have been observed in the emission lines \citep{middei_multi-wavelength_2018} or in the velocity components of the warm absorbers \citep{behar_multi-wavelength_2017}, suggesting the presence of a relatively distant outflow located several parsecs from the central engine \citep{peretz_multi-wavelength_2018}.

Photoionization modeling of the X-ray WAs has been performed using both XMM-Newton and Chandra spectroscopic data. All studies to date agree on the presence of a WA velocity component at approximately -600~km\,s$^{-1}$, first identified by \citet{blustin_multiwavelength_2003} and \citet{scott_intrinsic_2005}. A second, higher-velocity component at approximately -2000~km\,s$^{-1}$ has also been reported in more recent observations \citep{blustin_mass-energy_2007, behar_multi-wavelength_2017, peretz_multi-wavelength_2018, grafton-waters_multi-wavelength_2020}. Some authors find additional evidence for a third component around -900 km\,s$^{-1}$ \citep{behar_multi-wavelength_2017, grafton-waters_multi-wavelength_2020}, while one study suggests the presence of up to four velocity components ranging from –400 to –1800~km\,s$^{-1}$ in Chandra data \citep{mehdipour_multi-wavelength_2018}. Regarding emission line regions, velocity components near systemic velocity have been detected, as well as at –220 and –660~km\,s$^{-1}$, \citep{mehdipour_multi-wavelength_2018, grafton-waters_multi-wavelength_2020}.

While NGC~7469 has been observed frequently in the X-ray, the absence of X-ray integral field spectrographs (IFS) currently limits the ability to obtain spatially resolved velocity information in the X-ray regime. However, this galaxy has been observed by multiple optical integral field units (IFUs) \citep{Alonso-Herrero-2009, cazzoli_ngc_2020}, where evidence of an \ot $\lambda$5007\angstrom (\ot hereafter) outflow was detected with a maximum velocity of -715~km\,s$^{-1}$ \citep{robleto-orus_muse_2021, xu_spatially_2022}.

The present investigation focuses on tracing the extended soft X-ray emission using infrared (IR) IFU data and the aforementioned footprint lines. In addition to high spectral and spatial resolution, the wavelength range of the James Webb Space Telescope Mid-Infrared Instrument (JWST MIRI) and the Near-IR Spectrograph (NIRSpec) allow us to cover a wide range of emission lines, including those used in mid-IR diagnostic diagrams \citep{Feuillet2025} and IR footprint lines. To evaluate the effectiveness of the footprint lines in tracing soft X-ray emitting gas, we use both the XMM-Newton Reflection Grating Spectrograph (RGS) spectrum and Chandra imaging of NGC~7469. These datasets provide complementary high-resolution spectral and spatial information, enabling direct comparison with our IR IFU observations.

In addition to using footprint lines to track the very high ionization gas usually observed in the X-ray, we are also able to trace other phases of the outflow such as the high ionization gas and the warm ionized gas commonly observed in the optical/UV and optical respectively. To investigate these gas phases we use coronal lines (e.g. \nev $\lambda$14.32~\micns, \nev hereafter) and nebular lines (e.g. \sfo $\lambda$10.51~\micns, \sfo hereafter), allowing for direct comparison to the outflow parameters obtained from the footprint lines. This grants us the opportunity to determine which phase is the main dynamic driver of the outflow in NGC~7469.

The nuclear outflow has been detected in the IR, first in \cite{MullerSanchez2011}, in which the authors used the Very Large Telescope's Spectrograph for INtegral Field Observations in the Near Infrared (VLT/SINFONI IFU) to map the [Si \textsc{vi}] $\lambda$1.96~\mic emission line. This work was then extended to investigate the AGN outflow and its impact on the nuclear star formation using both the NIRSpec and MIRI IFUs \citep{u_goals-jwst_2022, lai_goals-jwst_2022, armus_goals-jwst_2023, Zhang2023, bianchin_goals-jwst_2024}. These investigations have highlighted the usefulness of IR lines in probing the nuclear environment of NGC~7469.

Despite these advances, the use of IR emission lines to trace soft X-ray gas, especially in terms of spatially resolved kinematics, has not been systemically explored. Our study addresses this gap using JWST’s unique capabilities in combination with archival X-ray data.

This paper is outlined as follows. Section \ref{sec:data} outlines the data used and data reduction processes. Section \ref{sec: footprint} demonstrates the effectiveness of using IR footprint lines to track the X-ray gas in NGC~7469. Section \ref{sec: global kinematics} demonstrates the global and spatially resolved kinematic information that may be obtained from IR lines compared to X-ray data. We calculate mass outflow rates and investigate the velocity structure of the X-ray gas in Section \ref{sec: calculations}. Finally, we discuss our findings and compare them with other nearby AGNs with strong X-ray outflows in Section \ref{sec: discussion}. We use Wilkinson Microwave Anisotropy Probe (WMAP) 9-year results cosmology throughout, with $H_0$ = 70.0 km s$^{-1}$ Mpc$^{-1}$ \citep{WMAP2013}.

\section{Observations and Data Reduction} \label{sec:data}

\subsection{MIRI and NIRSpec data}\label{sec: IR data}

The uncalibrated MIRI and NIRSpec data were retrieved from the Barbara A. Mikulski Archive for Space Telescopes (MAST). The MIRI and NIRSpec IFU observations of NGC~7469 were taken as part of a Director's Discretionary Time (1328; PIs: L. Armus, A. S. Evans) Early Release Science (ERS) program \citep{lai_goals-jwst_2022, u_goals-jwst_2022, armus_goals-jwst_2023}. 

The MIRI data \added{were taken using the Medium Resolution Spectrometer} and consist of spatially resolved spectra that range from 4.9 to 27.9 $\mu m$ over a field of view (FOV) up to $6\farcs6$ x $7\farcs7$ across four IFU channels. Each of the four IFU channels has three grating wavelength settings: short, medium, and long. The default FASTR1 readout pattern was used, with an exposure time of 444 seconds. Background observations were also conducted, as the galaxy is considered an extended source.

NGC~7469 was observed with NIRSpec in the high-resolution mode using the three available grating and filter combinations (G140H/F100LP, G235H/F170LP, and G395H/F290LP) corresponding to a wavelength range of 0.6 to 5.3 \micns, and used a large cycling four-point dither pattern to obtain a FOV of 4\farcs2 $\times$ 4\farcs8. The exposure time for each grating and filter combination is 817s. Once again, as NGC~7469 is considered an extended target, microshutter array leakage correction (leakcal) exposures were conducted to account for the leakage from the permanently open microshutters.

All of the footprint lines that we use are present in Channels 1 and 2 of the MIRI data, which have the best spatial resolution out of the MIRI channels. The pixel size of the NIRSpec data is smaller than the nominal pixel size of MIRI/ch1 and therefore needs to be convolved to a larger pixel size. We set values to the scalexy, ra\_center, and dec\_center parameters of the cube\_build step in both the MIRI and NIRSpec notebooks (\citealp{law_2025_notebooks}) to ensure matching cubes\footnote{We use values of 0\farcs196, 345.8151861, and 8.8739133 respectively for each parameter in both pipelines. We use pipeline versions v1.16.0 \citep{bushouse_2024_v16}, and v1.17.1 \citep{bushouse_2025_v17} for the MIRI and NIRSpec data respectively.}. The size of a spaxel is of 0\farcs196, which corresponds to 67~pc or a deprojected distance of 106~pc (r$_{dep}$ = r$_{obs}$/cos(i)) using a host inclination of 51$^{\circ}$ \citep{u_goals-jwst_2022}. However, despite the forced centering included in both the NIRSpec and MIRI pipelines, the data cubes have a small spatial offset with one another. This is a known issue (e.g. \citealp{Leftley2024}), which we address by centering our extractions around the peak in the continuum independently within each cube. 

\begin{figure}[]
\centering
\includegraphics[width=0.48\textwidth]{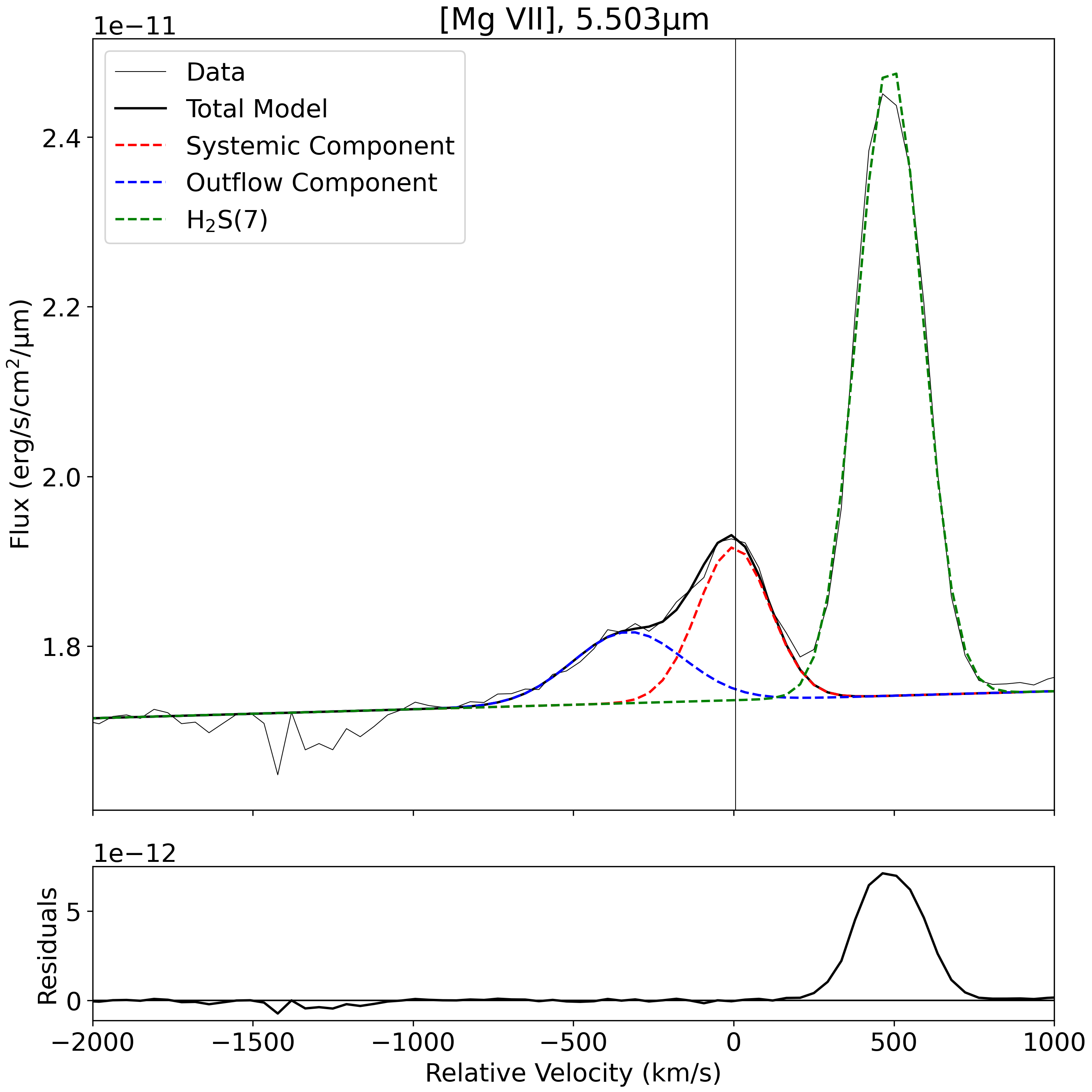}
    \caption{Multi-component spectral fit for the \mgsev $\lambda$5.50~\mic and adjacent H$_2$S(7) emission lines over the full spatial extent of the observed emission. As described in the text, each line in the data (gray) is modeled using two Gaussian components with systemic in red and the outflow component in blue. We also include a vertical gray line at systemic velocity.
    \label{fig: global line}}
\end{figure}

The emission line and continuum fitting procedure is similar to that described in \cite{Feuillet2025}. We first fit the \added{local} continuum emission \added{within a sufficiently large wavelength range centered on each emission line} and using a linear model\footnote{\nesix $\lambda$7.65~\mic is located on top of a strong polycyclic aromatic hydrocarbon (PAH) feature, and therefore requires a higher order polynomial to fit the continuum.}, then use \texttt{PySpecKit} \citep{Pyspeckit2011, Pyspeckit2022} to fit \added{each emission line independently} using \added{up to} two Gaussian components. \added{The routine uses a Levenberg-Marquardt optimizer to fit the emission lines, and will not return a second component if not necessary.} We consider the component with the smaller velocity shift to be the systemic component, and the blueshifted component to be the outflow component, and refer to it as such hereafter. In the case of \mgsev $\lambda$5.50~\micns, we also model the nearby H$_2$S(7) line to avoid any errors associated with it (see Figure \ref{fig: global line} for an example of our global line fitting). 

\subsection{XMM-Newton/RGS Data}\label{sec:RGS}

To enable a direct comparison between the IR footprint lines and soft X-ray emission, we require flux measurements for key X-ray lines. Values for several lines were previously measured using both older XMM-Newton RGS observations and Chandra High-Energy Transmission Grating Spectrometer/Medium Energy Grating (HETGS/MEG) data \citep{blustin_multiwavelength_2003, scott_intrinsic_2005, blustin_mass-energy_2007}. However, no previous study has reported fluxes for the \nitsix emission line, which is predicted by the global photoionization models that fit the newer, higher signal-to-noise RGS spectrum. As this line is among those used in our footprint line analysis, we reanalyze the most recent XMM-Newton RGS data to obtain consistent flux measurements for comparison with the IR lines.

In 2015, NGC~7469 was observed with XMM-Newton \citep{Jensen2001} seven times, one observation was carried out on June 13th, while the remaining six were performed between November 25th and December 30th. The exposure times ranged between 85~ksec and 101~ksec. We reprocessed and analyzed  each of the observations using the Science Analysis Software (SAS v21.1.0) and the standard software packages (\texttt{FTOOLS} v6.35.2, \texttt{XSPEC} v12.15.0, \citealp{Arnaud1996}). For the scientific analysis, we concentrated on the RGS data, which were  reduced using the standard SAS task {\it rgsproc} and the most recent calibration  files. We filtered for high background time intervals, applying a threshold of  0.2 counts s$^{-1}$ on the background event files.

The inspection of all seven RGS1 and RGS2 spectra revealed that four of the observations (December 15th, 23rd, 24th, and 26th 2015) caught NGC~7469 in a similar flux state. We thus combined the RGS1 and RGS2 of these four exposures in a single spectrum with the task {\it rgscombine}, which combines both the source spectra and the corresponding background and response matrices into a final single source, response and background file. The final combined RGS spectrum has a net exposure time of 596.8~ksec and a total of  316710  net counts (0.35--1.8 keV). The spectrum was then binned in constant wavelength bins of $\Delta \lambda\sim 0.03$~\angstromns, which is approximately half the Full-Width at Half-Maximum (FWHM) resolution of the instrument and the C-statistic was employed in all the fits.

We model the global RGS spectrum from 7-36\,\AA\, using power-law and blackbody continuum components, both absorbed by a neutral Galactic absorber. The effects of the ionized (warm) absorption has also been modeled, following the prescription in \cite{Behar2017} based on the 2015 RGS observations. Our values are consistent with the values found in that paper.

Gaussian emission components were then added to model the flux and any blue-shift of the \csixns, \nitsix and \osev emission lines. In the case of the \osev line, a broad underlying emission line, likely coming from a region close to the AGN, can be accounted for prior to modeling the narrow forbidden emission. In that case, the flux of the narrow component is of (3.6 $\pm$ 0.6) $\times$ 10$^{-14}$ erg\,cm$^{-2}$\,s$^{-1}$, which is about half of the previously reported fluxes. When fitting a Gaussian without the broad component, our results agree with previous measurements and proceed with this model for the remainder of the paper. The adjacent P-Cygni absorption is accounted for when modeling the \csix emission.

Based on the uncertainties on their fluxes reported in Table 1, the \osevns, \nitsixns, and \csix lines are detected at 6, 2 and 3$\sigma$ significance respectively. Accounting for the RGS line spread function, none of these narrow lines are resolved, thus their velocity widths are fixed at 0 and upper-limits are calculated on their FWHM values. These upper limits, as well as the measured value of the \osev line blueshift are also reported in Table \ref{tab: global kinematics}.



\begin{table}[h!]
\centering
\caption{Summary of the emission lines and best-fit fluxes}
\label{tab:lines}
\begin{tabular}{lccccc}
\hline
Line & $\lambda_{rest}$ & IP  & log(U) & Flux \\
&  & (eV) &  & 10$^{-14}$ erg cm$^{-2}$ s$^{-1}$ \\
\hline\hline
\multicolumn{5}{l}{XMM-Newton/RGS lines} \\
\hline
\osev & 22.10~\angstrom & 138  &  -0.25 & 6.3 $\pm$ 0.7 \\
\nitsix & 29.53~\angstrom & 98  &  -0.5 & 1.0 $\pm$ 0.5 \\
\csix & 33.74~\angstrom & 392  & 0.0  & 2.3 $\pm$ 0.7\\
\hline
\multicolumn{5}{l}{JWST/NIRSpec footprint lines} \\
\hline 
\sn  & 1.25~\mic    & 329 & 0.0  &  0.39 $\pm$ 0.12\\
\sisev  & 2.48~\mic & 205 & -0.75 & 2.06 $\pm$ 0.27 \\
\mgeig & 3.03~\mic  & 225 &  0.0 &  2.93 $\pm$ 0.33\\
\hline
\multicolumn{5}{l}{JWST/MIRI footprint, nebular and coronal lines}    \\
\hline
\feeig  & 5.45~\mic & 124 & -0.375  &  1.52 $\pm$ 0.29 \\
\mgsev  & 5.50~\mic     & 187 &  -0.375 &  1.67 $\pm$ 0.33 \\
\nesix  & 7.65~\mic & 126 &  -0.625  & 12.6 $\pm$ 1.72 \\
\sfo  & 10.51~\mic & 34.8 & -1.5  &   10.5 $\pm$ 1.10\\
\nev  & 14.32~\mic     & 97.1 & -0.875  &  13.5 $\pm$ 1.27 \\
\hline
\end{tabular}
\tablecomments{IP is the ionization potential of the ion in eV, while log(U) is the peak in ionization parameter (see Figure~\ref{fig: abund}).}
\end{table}

\subsection{Chandra/ACIS Data}\label{sec:chandra}

NGC~7469 was also observed with the Advanced CCD Imaging Spectrometer (ACIS) optimized for spectral resolving power with the High Energy Transmission Grating (HETG) on the Chandra X-ray observatory 10 times from 2002 to 2016. We obtained all available Chandra observations of NGC~7469 from the Chandra Data Archive, reprocessed and merged them using \href{https://cxc.cfa.harvard.edu/ciao/}{\texttt{CIAO}} (v4.17, \citealp{Fruscione2006}) with \texttt{CALDB} (v4.12.2). The total exposure time is 400~ksec. All Chandra datasets used in this work are publicly available through the Chandra Data Collection (CDC) at: \dataset[doi:10.25574/cdc.449]{https://doi.org/10.25574/cdc.449}.

We reprocessed all observations with the \href{https://cxc.cfa.harvard.edu/ciao/ahelp/chandra\_repro.html}{\texttt{chandra\_repro}} script, which applies the latest calibrations and enables sub-pixel resolution capabilities specific to ACIS-S. Individual datasets were aligned and merged using ObsID~2956, the longest single exposure, as the astrometric reference. Full-band (0.3–7~keV) images from each observation were visually inspected to check alignment quality. Centroid shifts between datasets were measured using the \texttt{CIAO} tool \href{https://cxc.cfa.harvard.edu/ciao/ahelp/dmstat.html}{\texttt{dmstat}}, applied to 0.3–7~keV images binned at 1/8 the native pixel size. Centroids were calculated within a $r=0\farcs2$ circular region centered on the brightest pixel. All observed centroid shifts were smaller than the native ACIS-S pixel scale of $0\farcs492$.

To ensure astrometric consistency across exposures, we cross-matched additional point sources and corrected any residual offsets using \href{https://cxc.cfa.harvard.edu/ciao/ahelp/wcs\_update.html}{\texttt{wcs\_update}}. Updated aspect solutions were incorporated using \href{https://cxc.cfa.harvard.edu/ciao/ahelp/reproject\_obs.html}{\texttt{reproject\_obs}}, and final merged datasets were generated with \href{https://cxc.cfa.harvard.edu/ciao/ahelp/merge\_obs.html}{\texttt{merge\_obs}}. 

To accurately model the Chandra point spread function (PSF), we simulated individual PSFs for each of the analyzed Chandra exposures. These simulations were carried out using \href{https://cxc.cfa.harvard.edu/ciao/PSFs/chart2/index.html}{\texttt{CHART}}, which accounts for the input source spectrum, off-axis angle, and aspect solution of each observation. The simulated ray traces were then projected onto the ACIS-S detector using \href{https://space.mit.edu/cxc/marx/}{\texttt{MARX}}, which incorporates parameters such as observation date, detector position, and exposure time. Each individual \texttt{CHART} simulation was converted into an event file after projection. These simulated PSF event files were then combined using \href{https://cxc.cfa.harvard.edu/ciao/ahelp/dmmerge.html}{\texttt{dmmerge}} to create a single, composite PSF suitable for use in our imaging analysis. All simulations were registered to a common astrometric frame using ObsID~29568 as the reference. The composite PSF was normalized and subtracted from data following the procedures detailed in \cite{Fabbiano2020}.

\begin{figure*}[]
\centering
\includegraphics[width=\textwidth]{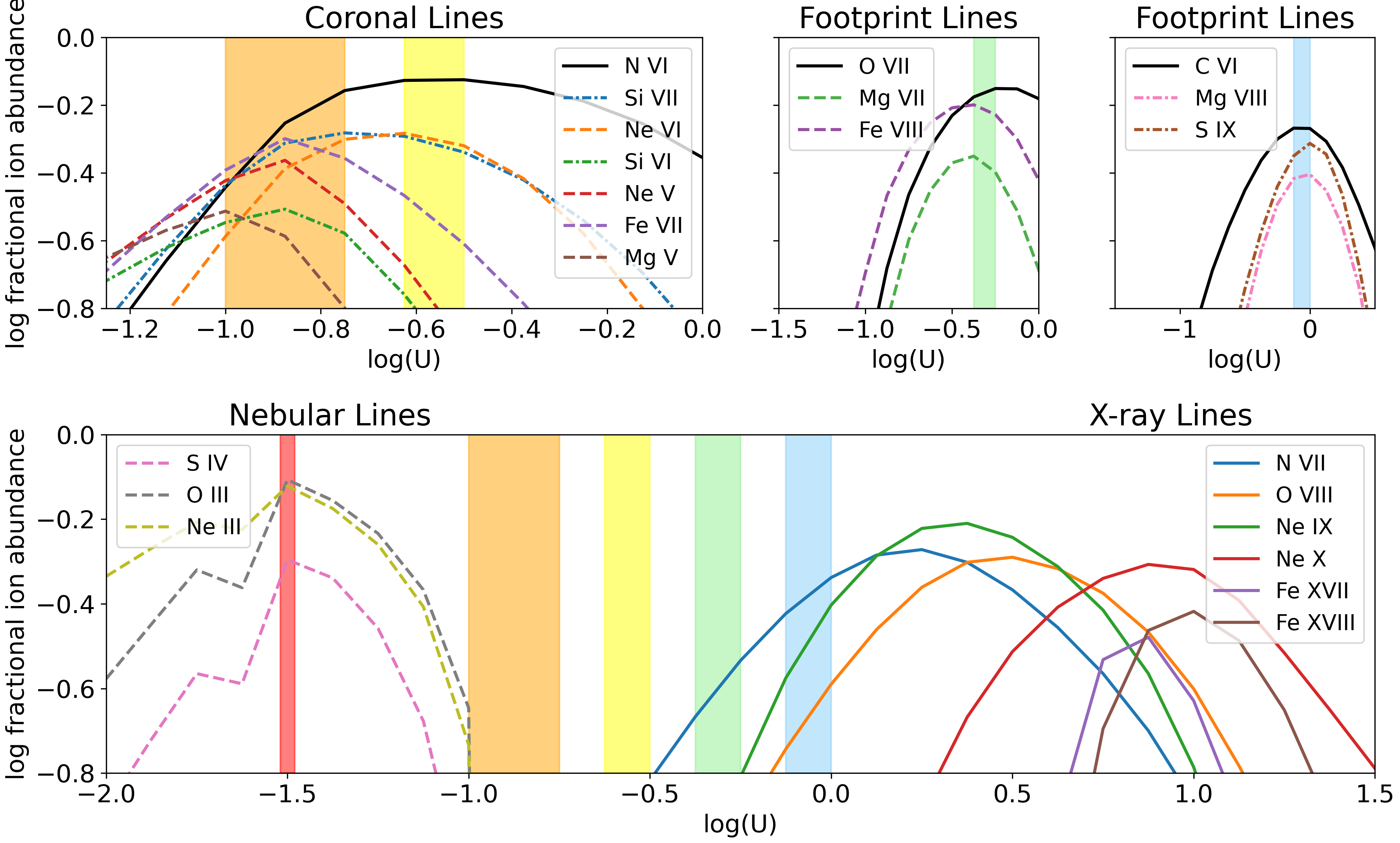}
    \caption{Cloudy predicted ionic abundances for the ions corresponding to all significantly detected emission lines in the X-ray, NIR and MIR in NGC~7469. The solid lines correspond to those found in the X-ray spectrum, while the dash-dotted and dashed lines correspond to those found in NIRSpec and MIRI, respectively. We separate the coronal and footprint lines (top panel) from the nebular and X-ray lines (bottom panel), and add bands emphasizing the range in ionization parameter where the different phases peak.
    \label{fig: abund}}
\end{figure*}

\section{Photoionization Modeling} \label{sec: footprint}

The ability of footprint lines to act as reliable tracers of X-ray gas lies not only in their IP values but also in the physical gas conditions from which they arise. Indeed, while the IP is a function of the atomic structure of the ion, the ionic abundances are dependent on the balance between ionization and recombination processes which are affected by the gas conditions such as the incident spectral energy distribution (SED).

Thus, comparing the range in ionization parameter (U) where the emission lines' corresponding ions peak in terms of ionic abundance is essential, as we cannot define footprint lines based solely on IP values. In \cite{Trindade2022} the authors used \texttt{Cloudy} models to determine that the \mbox{[Fe \textsc{x}]} $\lambda$6375~\angstrom line may be used to trace the X-ray gas in NGC~4151 as it was predicted to arise under the same gas conditions as the X-ray Ne \textsc{ix} (i) line. 

We used the photoionization code \texttt{Cloudy} (v23.01, \citealp{Cloudy2023}) to obtain the distribution of ionic abundances as a function of U for the ions corresponding to select detected emission lines in the X-ray, NIR, and MIRI spectra of NGC~7469 as shown in Figure~\ref{fig: abund}. We define U as
\begin{equation}\label{eq: U}
    U=\frac{Q}{4 \pi  r^{2} c n_{H}}
\end{equation}
where Q is the number of ionizing photons per second emitted by the central object (based on our SED, Q~=~$1.24\times10^{54}$~photons~s$^{-1}$), r is the distance between the central object and the cloud, c is the speed of light, and n$_{H}$ is the hydrogen number density \citep{Osterbrock2006}. We use the SED determined in \cite{Feuillet2025}, which is specifically tailored to NGC~7469 and included here for easy reference:
\begin{equation*}
    \alpha = -1.0 \textrm{ for } 1 \textrm{ eV}  \leq \textrm{h}\nu < 4.96 \textrm{ eV;}
\end{equation*}
\begin{equation*}
        \alpha = -0.92 \textrm{ for } 4.96 \textrm{ eV}  \leq \textrm{h}\nu \leq 13.6 \textrm{ eV;}
\end{equation*}
\begin{equation*}
        \alpha = -1.67 \textrm{ for } 13.6 \textrm{ eV}  \leq \textrm{h}\nu \leq 500 \textrm{ eV;}
\end{equation*}
\begin{equation*}
        \alpha = -0.80 \textrm{ for } 500 \textrm{ eV} \leq \textrm{h}\nu \leq 100 \textrm{ keV.}
\end{equation*}
For the models run to determine ionic abundances, we used similar parameters as those of \cite{Feuillet2025} which modeled the gas in NGC~7469 and use a column density of log N$_H$~=~21.5 cm$^{-3}$, Solar abundances\footnote{\added{Other abundances were later tested to model the measured emission lines but doing so did not improve the fit.}} \citep{Asplund2021}, and an arbitrary density of log n$_H$~=~4 cm$^{-3}$ as the ionic abundances are minimally dependent on n$_H$.


The corresponding peak in log(U) obtained from the models described above are demonstrated in Figure \ref{fig: abund} by the vertical bands and listed in Table \ref{tab:lines}. We define the nebular lines which peak at the same value of log(U)~=~-1.5 as that of the commonly used optical AGN warm outflow tracer \otns. Next, the ions corresponding to coronal lines all peak at a log(U)~=~-0.75 to -1. A interesting case is that of \nitsix which is an X-ray line but has an IP below that of 100~eV. \nesix $\lambda$7.65~\mic (\nesix hereafter) peaks at a similar log(U) and therefore traces the same gas phase, but we do not consider it to be in the same footprint line category as the following. The paired X-ray to footprint lines available to us are as such: 
\begin{itemize}[leftmargin=*]
    \item \osev~$\leftrightarrow$~\feeig~$\lambda$5.45~\micns,~\mgsev~$\lambda$5.50~\micns;
    \item \csix $\leftrightarrow$ \sn $\lambda$1.25~\micns, \mgeig $\lambda$3.03~\micns,
\end{itemize}
or \feeigns, \mgsevns, \sn, and \mgeig hereafter.

\begin{figure}[]
\centering
\includegraphics[width=0.45\textwidth]{mod_measured_luminosities.png}
    \caption{The relative luminosity of modeled values (L$_{mod}$/L$_{obs}$) is shown in brown circles. The closer the point is to unity (dashed black line), the better the modeled value agrees with the observed one. Only the observed IR luminosities and their uncertainties were used to determine the best model parameters, and the uncertainties are shown in green. The measured X-ray luminosity uncertainties are shown in pink.
    \label{fig: Modeled_luminosities}}
\end{figure}

As previously mentioned, the ability of a line to be considered a footprint line does not only depend on the IP. This is clearly seen here with [Si \textsc{vi}]~1.96~\micns, which has been investigated in the past using SINFONI \citep{MullerSanchez2011}. Indeed, although the line has a high IP of 167~eV the line does not peak within the same log~(U) range as any of the X-ray emission lines (bottom panel, Figure~\ref{fig: abund}). This suggests that the line cannot be used to track soft X-ray outflows in NGC~7469, but would track the same gas as the coronal line \nevns. 

After defining which lines correspond to the different gas phases, we created a grid of models to determine which combination reproduced our measured line luminosities best. We only used the IR lines for this fitting. Our values are well reproduced by a three-component model with components at log(U)~=~0,~-1.25,~\mbox{-2.25} (\texttt{footprint}, \texttt{coronal}, \texttt{nebular}) with respective column densities of log N$_H$~=~21.67, 21, and 20. \added{The agreement between our observed (L$_{obs}$) and modeled (L$_{mod}$) luminosities are shown in Figure \ref{fig: Modeled_luminosities}, where a relative luminosity (L$_{mod}$/L$_{obs}$) of 1 would indicate perfect agreement between the two values.} The three-component model reproduces the measured X-ray lines remarkably well, despite the large uncertainties resulting from the measurements.


\added{Our model reproduces all of the observed luminosities well within a factor of two, and most within reported uncertainties.} However, the \sn and \feeig luminosities are not as accurately reproduced by the photoionization model. This discrepancy in \snns, previously noted by \citet{Trindade2022}, may be due to uncertainties in the modeling of the collision strength for this transition or to an additional physical process not accounted for, such as photoexcitation or photofluorescence. The difference in the \feeig model could be due to depletion of iron onto dust grains. However, increasing the dust or depletion within the models would also affect the \mgeig and a selective depletion of Fe is unlikely. This issue was first raised in \cite{Ferguson1997}, in which they demonstrate that the Fe would need to be arbitrarily depleted by a factor of 10, but also point out that inaccurate model collision strengths could be the source of the issue. The inaccuracy of collision strength data for Fe$^{+7}$ is also pointed out in the reference paper for the atomic database used in \texttt{Cloudy} \citep{DelZanna2015}, and is therefore the more likely source of error.

\begin{figure*}[]
\centering
\includegraphics[width=\textwidth]{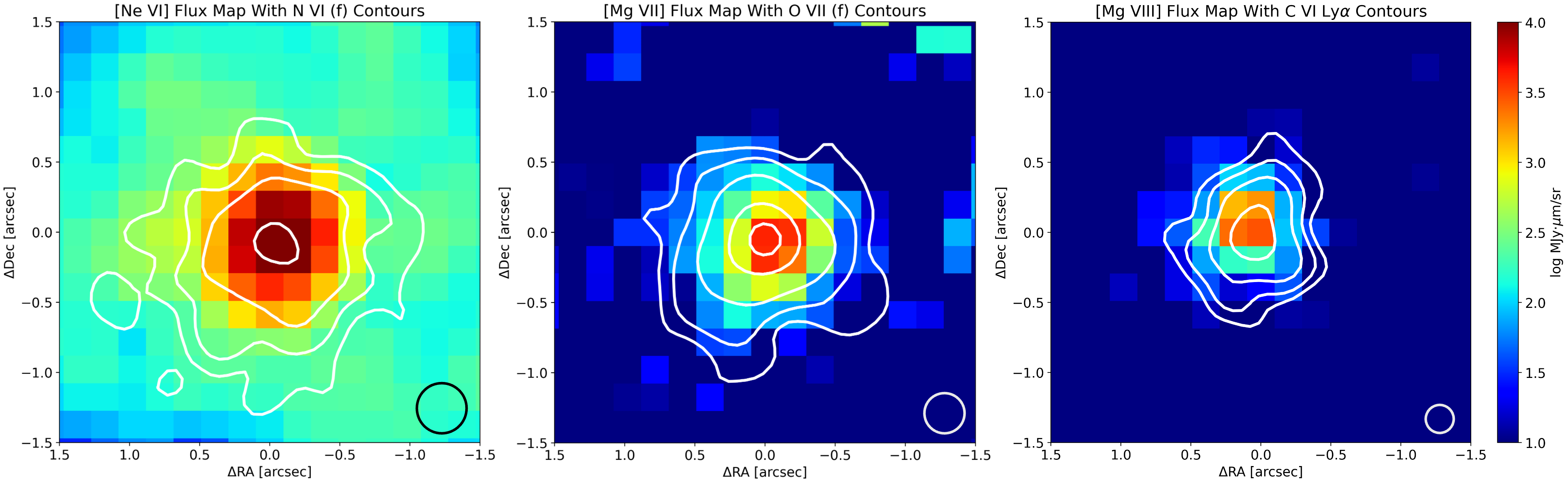}
    \caption{Comparing the morphology of the X-ray emission lines and the corresponding footprint lines. We find general agreement between them, with the caveats outlined in the text. The contour levels from outermost to innermost for \nitsixns, \osevns, and \csix are as follows: [0.15, 0.3, 0.6, 3], [0.35, 0.6, 1.2, 3, 5] and [0.22, 0.35, 0.6, 1.2] respectively and correspond to regions of equal count numbers. The Chandra images were PSF subtracted before creating the contours, as explained in Section \ref{sec:chandra}, and the PSF of the IR lines are shown as a circle in the lower right of each map\footnote{The MIRI PSF is calculated from the equation in \cite{Law_2023}. The NIRSpec PSF is approximated as $0\farcs2$ based on the models in \cite{Bentz_2025_NIRSpec_PSF} and the reported size of the nuclear region in \cite{bianchin_goals-jwst_2024}.}.
    \label{fig: chandra_comparison}}
\end{figure*}

\section{Morphology}

Using the Chandra images described in Section \ref{sec:chandra}, we examined the spatial extent and morphology of the soft X-ray emission in relation to the IR footprint lines outlined above. To isolate the relevant emission, we restricted the Chandra images to three narrow energy bands: 0.3–0.4~keV, 0.4–0.5~keV, and 0.5–0.6~keV, which correspond to \csixns, \nitsixns, and \osev respectively. The resulting images were then smoothed using a Gaussian kernel ($\sigma$~=~1.5) and set to a logarithmic scale. We then created contours based on these images to use in our comparison with the footprint lines. 

Each of the X-ray emission lines traced by the Chandra images has two corresponding IR footprint lines. However, consistent with our approach in analyzing spatially resolved parameters in the rest of this paper, we preferentially use the strongest of the two, which also corresponds to the line with the lower IP value, to trace the morphology in this section (\mgeigns, \mgsevns\footnote{The \mgsev and \feeig footprint lines have similar global flux values. However, as demonstrated in Section \ref{sec: footprint}, the models for \mgsev are better constrained than those for \feeigns. Therefore, we use \mgsev for the remainder of this investigation.}, \nesixns).

We use \texttt{QFitsView} \citep{Ott2012} to create our IR maps for the footprint lines, by integrating over the range of each emission line and removing the contribution from the continuum on each side. The resulting IR flux maps were then used for the base images, on which we overplotted the X-ray contours for direct visual comparison (Figure \ref{fig: chandra_comparison}).

There are multiple challenges that complicate the comparison between the soft X-ray and IR footprint morphologies. First, the X-ray emission lines in NGC~7469 are intrinsically faint, indeed, even with a very long exposure the \nitsix and \csix lines are still not considered to be statistically significant. Second, the Chandra images have not been continuum-subtracted, which signifies that the extracted bands include both line emission and direct, unattenuated X-ray continuum, which originates from the accretion disk/corona. In a Type 1 AGN such as NGC~7469, where the central engine is directly visible, the continuum contribution can be substantial and may blur or bias the apparent morphology. Finally, differences in spatial resolution and sensitivity between Chandra and JWST introduce additional uncertainties when comparing the extent and structure of the two datasets.

Despite these limitations, the soft X-ray morphology show a broad correspondence with that of the IR footprint lines. Both tracers exhibit extended emission with similar overall orientations, but we do detect footprint line emission in regions where the sensitivity of the X-ray images is too low. The general agreement between the two supports the interpretation that the footprint lines and soft X-rays are tracing overlapping phases of outflowing or photoionized gas.

\section{Global Kinematics} \label{sec: global kinematics}

\begin{table*}[]
\begin{center}
\caption{Global kinematic information for the X-ray emission lines and IR lines.}
\label{tab: global kinematics}
\begin{tabular}{cccccccccc}
\hline
Source & Instrument & Line  & IP & log(U) & v$_{05}$ &Component & FWHM$^{1, 2}$   & Velocity Shift$^{1, 2}$  &  Flux      \\
 & & & (eV) & & (km~s$^{-1}$) &     & (km~s$^{-1}$) & (km~s$^{-1}$) & (10$^{-14}$ erg s$^{-1}$ cm$^{-2}$)      \\
\hline\hline

(1)  & RGS   & \multirow{4}{*}{O \textsc{vii} (f)}  & \multirow{4}{*}{138} & \multirow{4}{*}{-0.25}     &  - &  -    & 0$^{+700}_{-0}$  & -400 $\pm$ 200 &   5.38 $\pm$ 1.5       \\
 (2)   & MEG &     & &   &  -  &   -   & 460 $\pm$ 220 & -200 $\pm$ 90   &  5.48 $\pm$ 2.25   \\
 (3)   & RGS     &    &   &   &  -   &   -   & 1200$^{+1000}_{-700}$ & –520 $\pm$ 200 & 6.29 $\pm$ 0.90 \\
  (4)   & RGS     &    &   &   &  -   &   -   & $<$450 & -420 $\pm$ 110 & 6.3 $\pm$ 0.7 \\
 \hline
 (1) &  \multirow{3}{*}{RGS}  & \multirow{3}{*}{C \textsc{vi} Ly$\alpha$}      & \multirow{3}{*}{392}   & \multirow{3}{*}{0}  &  -  &   -   & 700$^{+800}_{-700}$   & 700 $\pm$ 200  &     2.12 $\pm$ 0.76      \\
(3)   &     &    &   &  &  -  &   -   & 0$^{+2000}_{-0}$      & -70$^{+240}_{-470}$              &     1.76 $\pm$ 0.59       \\
(4)   &     &    &   &   &  -   &   -   & $<$1150  & $<$180 & 2.3 $\pm$ 0.7  \\
\hline\hline

 &   \multirow{6}{*}{NIRSpec}        & \multirow{2}{*}{[Mg \textsc{viii}]} & \multirow{2}{*}{225}& \multirow{2}{*}{0}  & \multirow{2}{*}{-789}& Sys  &  289    &    -52     & - \\
 &          &                                     &               &      &  & Out  &   694   &  -508       & - \\
 
 &          & \multirow{2}{*}{[S \textsc{ix}]}   & \multirow{2}{*}{329}  & \multirow{2}{*}{0} & \multirow{2}{*}{-865} & Sys  &   237   &    -1     & - \\
 &          &                                     &              &       &                                              & Out  &  777    &    -560     & - \\
  &  & \multirow{2}{*}{[Si \textsc{vii}]}  & \multirow{2}{*}{205} & \multirow{2}{*}{-0.75} & \multirow{2}{*}{-1018} & Sys  &  313    &   -56      & - \\
 &                          &                               &      &                    &                           & Out  &   777   &    -764     & - \\
\hline

 & \multirow{10}{*}{MIRI} & \multirow{2}{*}{[Fe \textsc{viii}]} & \multirow{2}{*}{124} &  \multirow{2}{*}{-0.375} & \multirow{2}{*}{-718}       & Sys  &  347   &    26     & - \\
 &          &                            &         &                               &                                    & Out  &   695    &    -429     & - \\

 &         & \multirow{2}{*}{[Mg \textsc{vii}]}  & \multirow{2}{*}{187} & \multirow{2}{*}{-0.375} & \multirow{2}{*}{-532}   & Sys  &  253    &     0    & - \\
 &          &                        &             &                      &                                               & Out  &   399  &    -424     & - \\
  &        & \multirow{2}{*}{[Ne \textsc{vi}]}   & \multirow{2}{*}{126} & \multirow{2}{*}{-0.625}  & \multirow{2}{*}{-598}   & Sys  &   245   &   -69      & - \\
 &          &                            &         &                     &   & Out  &    504   &    -502     & - \\
   &         & \multirow{2}{*}{[Ne \textsc{v}]}  & \multirow{2}{*}{97.1} & \multirow{2}{*}{-0.875} & \multirow{2}{*}{-736}   & Sys  &  266    &     -55    & - \\
 &          &                        &             &                      &                                               & Out  &   710  &    -505     & - \\
  &         & \multirow{2}{*}{[S \textsc{iv}]}  & \multirow{2}{*}{34.8} & \multirow{2}{*}{-1.5} & \multirow{2}{*}{-630}   & Sys  &  301    &     -59    & - \\
 &          &                        &             &                      &                                               & Out  &   521  &    -621     & - \\
 \hline
\end{tabular}
\tablecomments{(1) \cite{blustin_multiwavelength_2003}, (2) \cite{scott_intrinsic_2005}, (3) \cite{blustin_mass-energy_2007}, (4) this paper. The previous papers use an older value for the redshift of NGC~7469 of z = 0.01639. This discrepancy corresponds to values being greater by $\sim$35~km~s$^{-1}$. $^1$ The IR FWHM values are corrected for instrumental broadening and the velocity shifts are deprojected as described in the text. $^2$ The FWHM and velocity shift values do not include an uncertainty as the errors output by the fitting routine were insignificant (on the order of 1 to 10 km\,s$^{-1}$) \added{compared to the spectral resolution of the instruments}. \added{Indeed, while the velocity shifts output by the fitting routine are not exactly 0, they are indistinguishable from zero again due to the resolution of the instruments.}}
\end{center}
\end{table*}

Having demonstrated our ability to use footprint lines to track the X-ray gas in NGC~7469, we compared the global kinematics of the IR and X-ray emission lines. The X-ray data are not high enough quality to extract any kinematic information for the \nitsix line, but values for the FWHM and velocity shift were reported for the \osev and \csix lines \citep{blustin_multiwavelength_2003, scott_intrinsic_2005}. 

\begin{figure}[h!]
\centering
\includegraphics[width=0.45\textwidth]{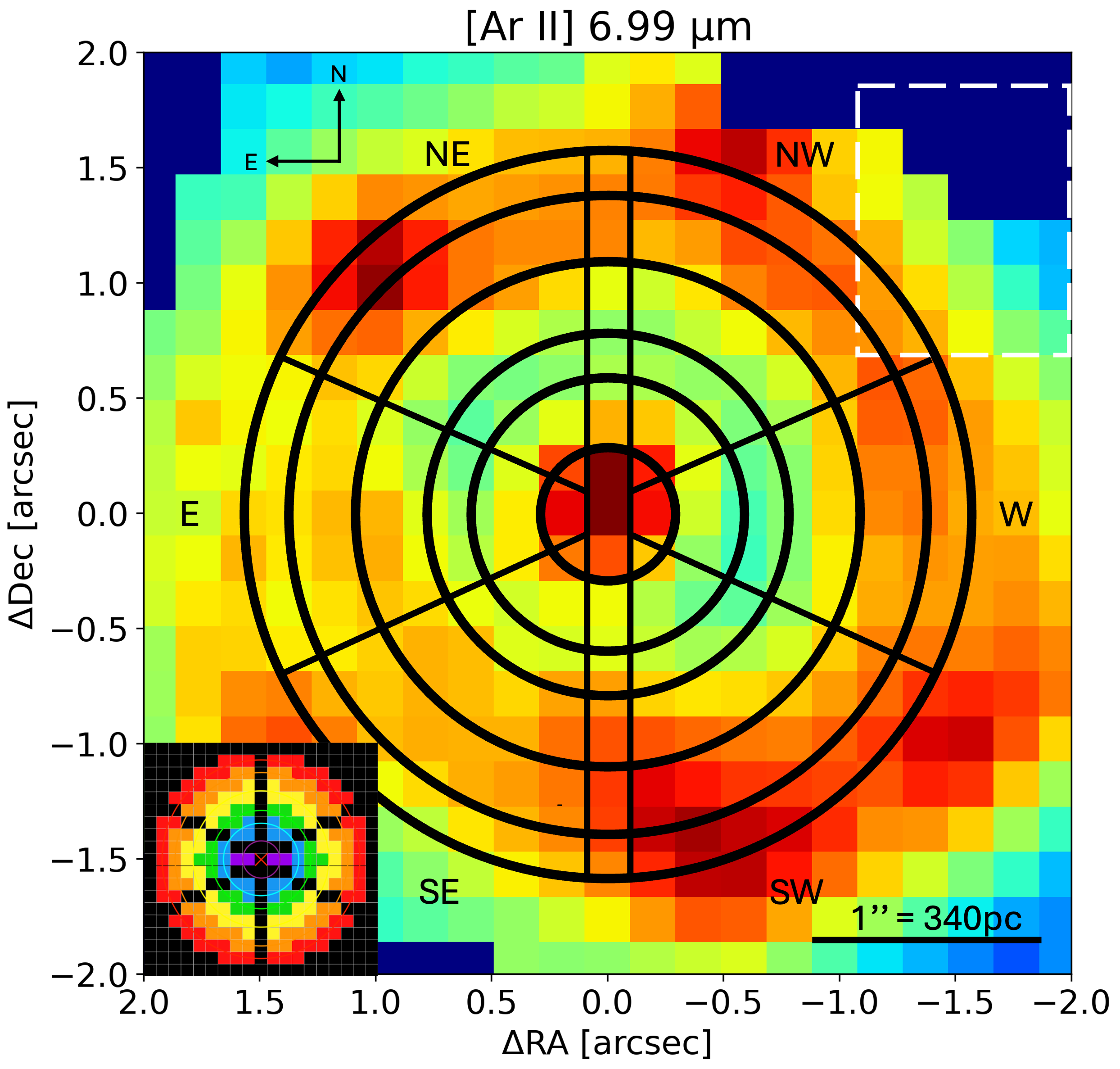}
    \caption{We show a continuum-subtracted map of the [Ar~\textsc{ii}]~$\lambda$6.99~\mic line, a star-formation indicator in MIRI/ch1 to demonstrate the position of our regions with respect to the star-forming ring. The map was created using \texttt{QFitsView} \citep{Ott2012}. We over-plot the average distance from the center for each circular region of interest as well as the way the circles were further split into six directions: E, NE, NW, W, SW, and SE. The actual spaxels making up each region are shown in the lower left inlay. The white dashed region is the redshifted outflow region detailed in Appendix \ref{app: redshifted outflow}.}
    \label{fig: regions}
\end{figure}

The global flux values were obtained from fitting the emission line spectrum using all spaxels within the entire grid region shown in Figure~\ref{fig: regions}, using the method described in Section~\ref{sec: IR data}. For each emission line, we report the values for the FWHM and the velocity shift of each component in Table \ref{tab: global kinematics}. The FWHMs are corrected for instrumental broadening \mbox{($\sigma_{cor}$ = $\sqrt{\sigma_{uncor}^2-\sigma_{inst}^2}$)}. The instrumental FWHM values are calculated from the resolving power values reported in \cite{Argyriou2023} for MIRI and \cite{Jokobsen2022} for NIRSpec. 

The velocity shift values also need to be deprojected due to the geometry of the system. To do so, both the inclination of the bicone \added{(assuming a biconical geometry)} and the inclination of the galaxy should be taken into account. However, this galaxy was not fit by \cite{Fischer2013} due to the compactness of the emission, \added{therefore we do not have any geometrical constraints on the outflow (i.e. opening angle or inclination)}. We do not believe that the cone is fully directed into the plane of the galaxy based on the weak coupling between the outflow and the star-forming ring \citep{lai_goals-jwst_2022, Feuillet2025}. Therefore, we only deproject the velocities by \mbox{v$_{dep}$ = v$_{obs}$/sin(i)} or 29\% when using the inclination of the host of 51$^{\circ}$ \citep{u_goals-jwst_2022}. The resulting velocities are therefore minimum values, and a 45$^{\circ}$ bicone angle would increase the velocities by an additional 41\%. 

We also define the maximum outflow velocity, v$_{max}$, as the velocity at which a fixed fraction of the total modeled line flux is enclosed:
\begin{equation}
    F(v_{max}) = \frac{\int_{-\infty}^{v_{max}}f(v)dv}{\int_{-\infty}^{\infty}f(v)dv}
\end{equation}
Common choices for this fraction are 2\% (e.g., \citealp{Harrison2014_02, Bessiere2022_02}) or 5\% (e.g., \citealp{Zamaska2014_05, Rose2018_05}). Following \cite{Rose2018_05}, we adopt the 5\% criterion (v$_{05}$), as it reduces sensitivity to uncertainties in continuum placement. The maximum outflow velocity values are \added{not corrected for inclination like the velocity shifts were}, and are also detailed in Table \ref{tab: global kinematics}.

The range of FWHM values for \osev is consistently reported across all three papers as 500–680 km~s$^{-1}$, while the range for \csix extends from 0 to 1500 km~s$^{-1}$. Clearly, the FWHM measurements from the X-ray data are highly uncertain, and the velocity shifts of \csix cannot be considered reliable due to the presence of an absorption component in its P-Cygni profile. Perhaps best to say that the FWHM values for the X-ray lines are mainly upper limits, e.g. consistent with less than a few 100 km/s for O VII and less well constrained for C VI.

In contrast, the velocity shift of \osev is more consistent across the three studies, with values clustering around –400 km~s$^{-1}$. This is remarkably similar to the velocity shifts of \osevns's corresponding IR footprint lines, measured at –424 and –429 km~s$^{-1}$. The new RGS measurement of \osev is also consistent, with a velocity of $-420~\pm~110$\,km\,s$^{-1}$. The overall agreement between the kinematic information derived from the X-ray lines and the IR footprint lines reinforces their reliability as tracers of soft X-ray emission.


\section{Calculations} \label{sec: calculations}

\subsection{Mass of the ionized gas}\label{sec: masses}

The calculation of gas masses for a defined region ($M_{reg}$) is conducted by using both the measured line luminosity and \texttt{Cloudy} model predicted fluxes for the emission line using the following equation: 

\begin{equation}\label{eq: M}
    M_{reg} = N_H \mu m_p \left( \frac{L_{obs}}{F_{mod}} \right)
\end{equation}
where $N_H$ is the column density, $\mu$ is 1.3, $m_p$ is the mass of a proton, $L_{obs}$ is the observed luminosity of the emission line\footnote{We do not correct the luminosities for reddening, consistent with other JWST investigations into this target.}, and $F_{mod}$ is the modeled flux of the emission lines according to our \texttt{Cloudy} model. 

The above equation is the standard reported equation for investigations of this sort, but a more detailed explanation is given here. Our models are physically constrained by two parameters: they must not exceed the column of gas available in that particular region (thickness, $\delta$r) or the maximum emitting area at that distance. For each measured line luminosity we solve the following equation:
\begin{equation}\label{eq: areas_expl}
    L_{obs} = F_{\texttt{f}} \times A_1 + F_{\texttt{c}} \times A_2 + F_{\texttt{n}} \times A_3
\end{equation}
where F$_{\texttt{f}}$ is the flux from the \texttt{footprint} \texttt{Cloudy} model and A$_1$ is the corresponding emitting area, similarly, F$_{\texttt{c}}$ and F$_{\texttt{n}}$ are the fluxes from the \texttt{coronal} and \texttt{nebular} models, A$_2$ and A$_3$ are the respective emitting areas. Once we have solved for A$_1$, A$_2$, and A$_3$, ensuring that the areas do not exceed the maximum value, the mass is then obtained with:
\begin{equation*}
    M_{\texttt{f}} = A_1 n_{H_{\texttt{f}}} \delta_{\texttt{f}} \mu m_p
\end{equation*}
\begin{equation*}
    M_{\texttt{c}} = A_2 n_{H_{\texttt{c}}} \delta_{\texttt{c}} \mu m_p
\end{equation*}
\begin{equation*}
    M_{\texttt{n}} = A_3 n_{H_{\texttt{n}}} \delta_{\texttt{n}} \mu m_p
\end{equation*}
\begin{equation*}\label{eq: M2}
    M_{reg} = M_{\texttt{f}} + M_{\texttt{c}} + M_{\texttt{n}}  \\
\end{equation*}

In this case, the total mass of the ionized gas is calculated using the three component \texttt{Cloudy} models described in the previous section. We also use an average deprojected distance and Equation \ref{eq: U} to calculate the density. 

To obtain the total mass from our measurements and models, we used a constrained weighted least squares method to solve our eight equations with 2 unknowns (Equation \ref{eq: areas_expl}) and with different levels of uncertainty. This method takes into account the reliability of each measurement by weighting the ones with smaller uncertainties more. The best-fit solution and their uncertainties were then calculated using matrix operations. The resulting total mass of the gas across the three phases is (4.05 $\pm$ 0.39) $\times$ 10$^{6}$ M$_\odot$.

Thanks to the spatial resolution of the NIRSpec and MIRI instrument, we can obtain spatially resolved mass values in addition to the global value. To do so we define and use the smaller regions again shown in Figure \ref{fig: regions} and compute \texttt{Cloudy} models for each (the parameters for each region are reported in Appendix \ref{app: mod_params}). The regions become progressively larger further from the nucleus in order to improve S/N in the resulting combined spectra. We ensure once again that the maximum thickness and area of each region is not exceeded. For the mass calculations and the following spatially resolved outflow parameters, we focus our attention on a characteristic line of each ionization regime: \mgeig for the \texttt{footprint} component, \nesix for the \texttt{coronal} component, and \sfo for the \texttt{nebular} component. These lines were both the strongest and best reproduced by the global \texttt{Cloudy} models within the lower MIRI channels.

\begin{figure*}[]
\centering
\includegraphics[width=\textwidth]{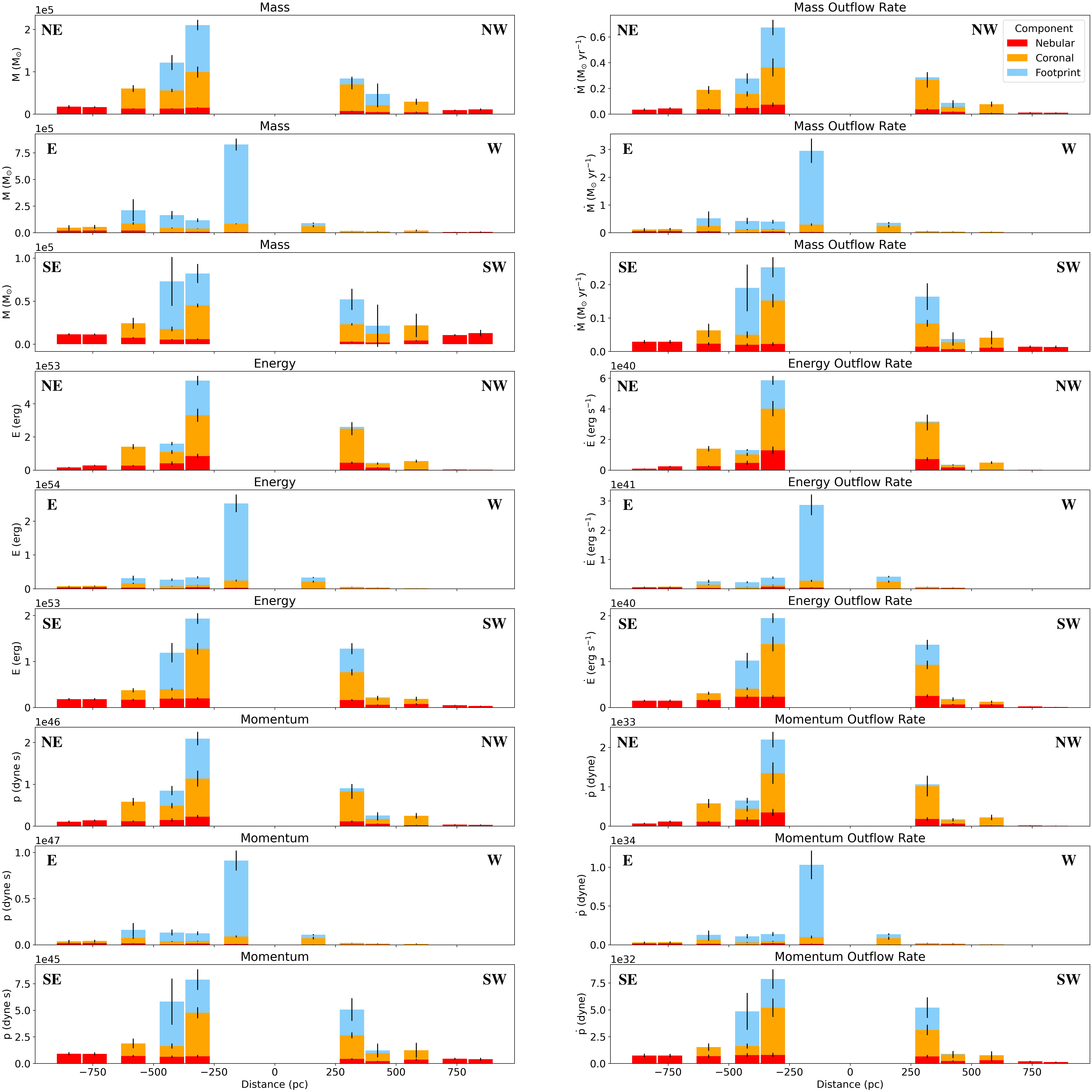}
    \caption{Spatially resolved calculated outflow parameters. We show the distribution of the various parameters obtained by reporting the values obtained from each region as defined in Figure \ref{fig: regions}. The negative distances correspond to the East side regions, while the positive distances correspond to the West. We split each parameter between the North, East-West, and South ``cones" once again defined in Figure \ref{fig: regions}. Finally, we split our values between the nebular, coronal, and footprint components as described in the text.}
    \label{fig: outflow_params_hist}
\end{figure*}

\subsection{Outflow Parameters} \label{sec: outflow parameters}

The kinematic information reported in Section \ref{sec: global kinematics} corresponds to global fits to the emission lines, however, once again thanks to the spectral resolution of the JWST IFU instruments, we are able to extract spatially resolved kinematic information. This allows us to better constrain the impact of the outflow on the surrounding medium. Helpful measurements include calculations of the mass outflow rate ($\dot{M}$), kinetic energy, and momenta. We calculate the mass outflow rate using:
\begin{equation}
    \dot{M} = \frac{M_{out}v}{\delta r}
\end{equation}
where M is the mass in each region, v is the maximum deprojected velocity corrected for inclination, and $\delta$r is the deprojected size for each extraction\footnote{We use the thickness value as the extraction size, $\delta$r, as was defined in the previous section.}.

We also calculate the kinetic energy (E), kinetic energy flow rate ($\dot{E}$), momentum (p), and momentum flow rate ($\dot{p}$) are given by:

\begin{equation}
E = \frac{1}{2}M_{out}v^2,
\end{equation}
\begin{equation}
\dot{E} = \frac{1}{2}\dot{M}_{out}v^2,
\end{equation}
\begin{equation}
p = M_{out}v,
\end{equation}
\begin{equation}\label{eq: pdot}
\dot{p} = \dot{M}_{out}v.
\end{equation}

We show the spatially resolved outflow parameters for each region in Figure \ref{fig: outflow_params_hist}. The negative distances correspond to the East side, while the positive distances reflect the West side values. The values are further divided as was shown in Figure \ref{fig: regions} between the North, E-W and South. We finally separate the results for these parameters separately for each of the three ionization components. 

\begin{figure*}[]
\centering
\includegraphics[width=\textwidth]{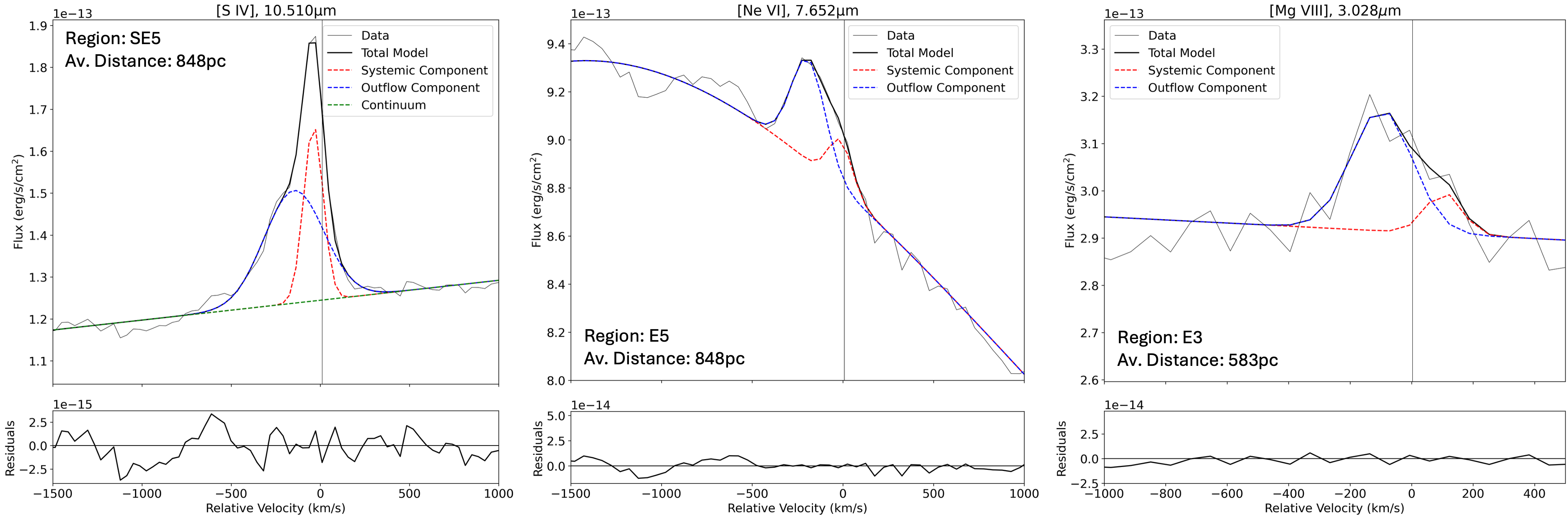}
    \caption{Example spectra and fitted emission lines for the furthest regions where the respective emission lines are detected: \sfo and \nesix are detected up to 848 pc from the nucleus, while \mgeig is detected up to 583 pc, as explained in the text.}
    \label{fig: spatial_lines}
\end{figure*}

\added{Note that all three components are not present in every region. This is due to the emission being too weak to fit an emission line at larger distances for the higher ionization ones. Specifically, the \sfo line is detected in every region up to a distance of 848\,pc, \nesix is detected to 848\,pc in the E region and 583 pc in all other regions, while \mgeig is detected to 583\,pc in the E region and to varying extents in the other regions. We show an example of the fitted emission lines within the regions at the largest distance from the nucleus in Figure \ref{fig: spatial_lines}.} We can draw two main conclusions from the distributions in Figure \ref{fig: outflow_params_hist}: the parameters are higher for the East side of the nucleus and are higher for the \texttt{footprint} component when present.

We report numerical values in Table \ref{tab: outflow parameters details}, where outflow rates are given in terms of maximum value rather than total, as we are interested in the location of the maximum deposited mass, energy, and momentum. We include integrated values for the mass, the energy, and the momentum. Finally, we report the $\dot{E}_{max}/L_{bol}$ parameter using a bolometric luminosity value of L$_{bol}$~=~(2.84~$\pm$~1.41)\,$\times$\,10$^{44}$~erg~s$^{-1}$, obtained from averaging the reported values in \cite{Feuillet2025} .

\section{Implications for AGN Feedback} \label{sec: discussion}

\begin{table*}[]
\begin{center}
\caption{Summary of calculated outflow parameters for the three gas phase components}
\label{tab: outflow parameters details}
\begin{tabular}{cccccccc}
\hline
Component & Total M  & $\dot{M}_{max}$ & Total E & $\dot{E}_{max}$ & Total p & $\dot{p}_{max}$ & $\dot{E}_{max}/L_{bol}$ \\
 & 10$^5$ M$_{\odot}$ & M$_{\odot} $yr$^{-1}$ & 10$^{53}$ erg & 10$^{40}$ erg s$^{-1}$ & 10$^{47}$ dyne s & 10$^{33}$ dyne  & \% \\
& (1) & (2) & (3) & (4) & (5) & (6) & (7) \\
\hline\hline
\texttt{Nebular} & 2.85 $\pm$ 0.09 & 0.073 $\pm$ 0.015 & 6.24 $\pm$ 0.25 & 1.29 $\pm$ 0.24 & 0.24 $\pm$ 0.01 & 0.35 $\pm$ 0.09 & 0.0046 $\pm$ 0.0028\\
\texttt{Coronal} & 7.78 $\pm$ 0.44 & 0.29 $\pm$ 0.07 & 17.64 $\pm$ 0.78 & 2.72 $\pm$ 0.50 & 0.73 $\pm$ 0.04 & 1.00 $\pm$ 0.27 & 0.0096 $\pm$ 0.0058 \\
\texttt{Footprint} & 14.4 $\pm$ 1.35 & 2.66 $\pm$ 0.44 & 34.56 $\pm$ 2.78 & 25.9	$\pm$ 3.52 & 1.37 $\pm$ 0.14 & 9.31	$\pm$ 1.84 & 0.091 $\pm$ 0.054 \\
\hline
\end{tabular}
\tablecomments{Columns are (1)~gas mass, (2)~mass outflow rate, (3)~kinetic energy, (4)~kinetic energy flow rate, (5)~momentum, (6)~momentum flow rate, and (7)~efficient feedback parameter. These values are all obtained from Equations \ref{eq: M}-\ref{eq: pdot}.}
\end{center}
\end{table*}

\subsection{Contextualizing Results with Nearby Seyferts}

\begin{figure*}[]
\centering
\includegraphics[width=\textwidth]{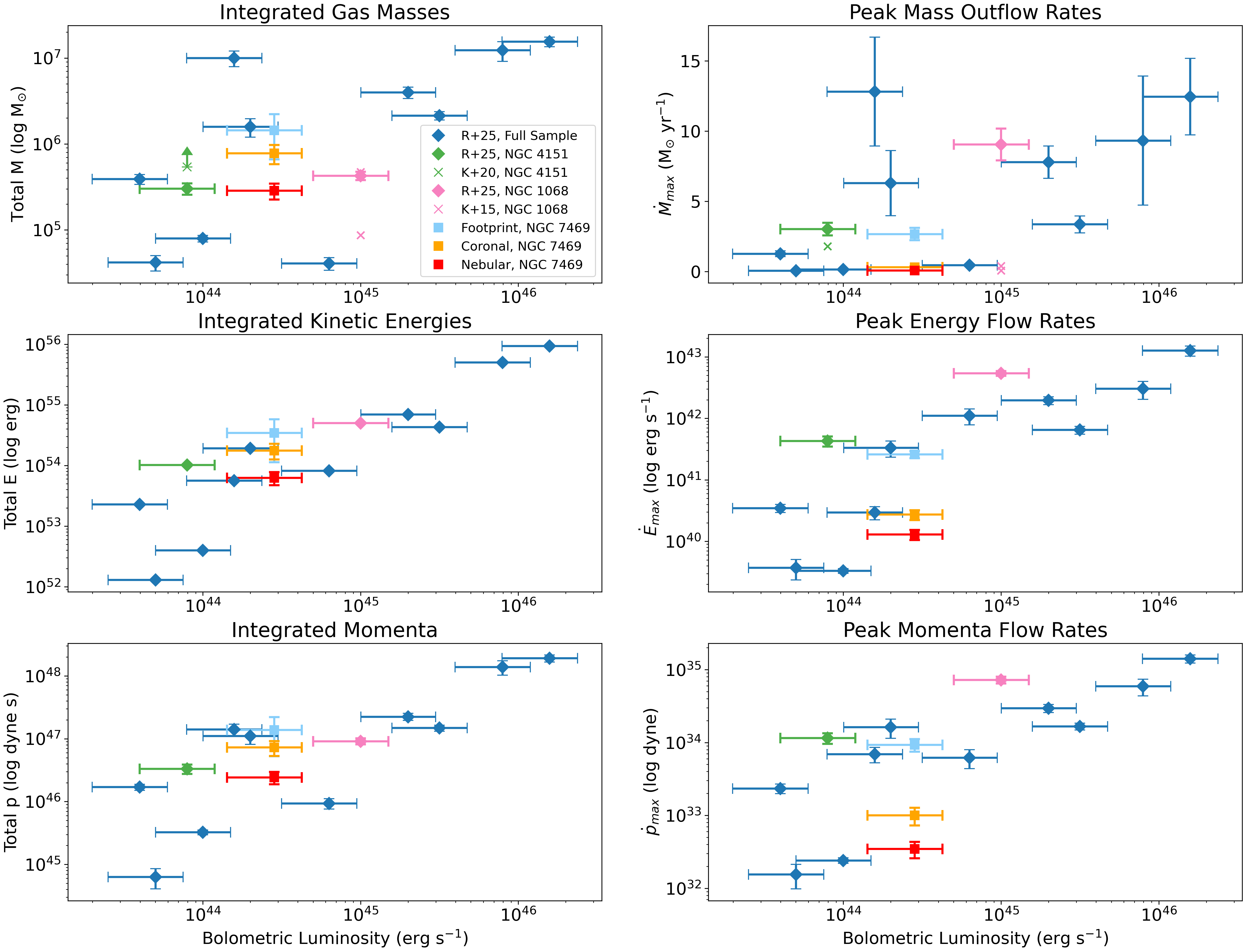}
    \caption{We compare the outflow parameters calculated in this paper for the nearby Seyfert galaxy NGC~7469 (squares) split into the three ionization components: footprint (light blue), coronal (orange), nebular (red), to the sample of nearby Seyfert galaxies in \cite{Revalski2025} (R+25, dark blue). We highlight two objects, NGC~4151 (green; \citealp{Crenshaw2015}) and NGC~1068 (pink) which are the two AGN with high efficient feedback parameters (see Figure \ref{fig: feedback_param}). We include the mass and mass outflow rates calculated for these two galaxies using X-ray data in \cite{Kraemer2015} and \cite{Kraemer2020} (K+15, K+20, $\times$-markers).
    \label{fig: outflow_params_compare}}
\end{figure*}


To put the outflow parameters measured in this paper into context, we compare our results to those measured in \cite{Revalski2021} and \cite{Revalski2025} (see Figure \ref{fig: feedback_param}). In their papers, the authors studied AGN feedback in 12 nearby (z $<$ 0.05) Seyfert galaxies. All of the galaxies in the sample are Seyfert 2s with the exception of NGC~4151, which is a Type 1. These feedback parameters were derived using the optical \ot line, which is a commonly used tracer of AGN activity, but arises from gas in a different ionization state than that analyzed using the footprint line method, which traces soft X-ray gas. 


To compare the differences between mass and mass outflow rate measurements in the optical and X-ray, we compare the \ot based values for NGC~4151 and NGC~1068 to those derived by \cite{Kraemer2020} and \cite{Kraemer2015} using X-ray emission line spectra. As described in \cite{Revalski2018}, when lacking spatial resolution, which was the case for the X-ray papers, the choice of filling factor and distance from the nucleus can affect the final measurements by several orders of magnitude. While the optical and X-ray values for NGC~4151 agree reasonably well, the values for NGC~1068 differ significantly. The discrepancy is likely due to the choice of X-ray filling factor, but the fact that these studies track gas in different ionization states gas may also contribute to the differences. These issues highlight the significance and importance of our analysis which provides a self-consistent approach to analyzing the different phases of the outflow.


Our measured values for the masses and outflow parameters in NGC~7469 are generally consistent with the trends observed in nearby Seyfert galaxies across all three components. Notably, the highest values are associated with the X-ray gas component, largely due to its higher derived mass relative to the \texttt{nebular} component. Similar results are also reported in \cite{falcão2025hotphotoionizedxraygas}.

\subsection{Comparison with Other NGC~7469 Outflow Parameter Measurements}


The outflow parameters for NGC~7469 have been calculated in the past using a variety of methods and data. The present paper represents the first spatially resolved investigation into these parameters, which has been shown to deliver the most accurate results \citep{Revalski2018}. 

The kinetic energy outflow rate for the X-ray gas has been calculated using X-ray absorbers in \cite{crenshaw_feedback_2012}, \cite{blustin_mass-energy_2007}, and \cite{peretz_multi-wavelength_2018}. The resulting values range from 1.2\,$\times$\,10$^{40}$ to 1.1\,$\times$\,10$^{43}$ erg\,s$^{-1}$, with our footprint phase values falling within this range\footnote{The UV-derived values range from 1.0\,$\times$\,10$^{39}$ to 4.1\,$\times$\,10$^{41}$ erg\,s$^{-1}$, which encompasses our result for the coronal gas phase.}. As evidence by the three orders of magnitude spread in energy flow rates previously calculated only by tracing the X-ray gas, obtaining global values for this parameter is highly uncertain and once again extremely sensitive to the choice of filling factor and distance from the AGN. Obtaining spatially resolved measurements removes these issues from the process, offering better constrained values as well as spatial information on the feedback effects. 



We can compare the \texttt{nebular} gas phase values to those derived using the \ot line and MUSE IFU data in \cite{xu_spatially_2022}, as they were shown to arise within the same gas (see Figure \ref{fig: abund}). It is important to note that the authors in \cite{xu_spatially_2022} use different equations and methods from those used in the current paper, and, once again, fit only a global value rather than providing spatially resolved fits (except for the separation of AGN and SF regions). They find kinetic energy flow rates of 3.2 and 5.5 $\times$ 10$^{41}$ erg s$^{-1}$ for the SF and AGN regions, and corresponding mass outflow rates of 3.9 and 6.8 M$_{\odot} $yr$^{-1}$, which are about one order of magnitude greater than the values we derived from the nebular gas phase.  


Our results paint a picture of the multiphase outflow in NGC~7469 where the very high-ionization X-ray component dominates the kinetic energy outflow rate (approximately 20 times greater than the \texttt{nebular} component), which is directly related to the efficiency of feedback processes. This is consistent with the findings of \cite{Trindade2021_xray_outflows}, where the authors found that in Mrk 34, the X-ray gas provides a kinetic energy outflow rate an order of magnitude greater than that of the \ot gas. 

\subsection{Feedback Efficiency in NGC 7469}


The feedback processes in NGC~7469 have been extensively studied in the past using a variety of techniques and indicators. One of the more recent studies involved the investigation of PAH and molecular hydrogen in the central region of this galaxy. The authors found that the H$_2$/PAH ratios were consistent with star formation and detected minimal impact of the AGN on the ring. Similarly, in \cite{Feuillet2025}, we suggested a scenario of weak coupling between the AGN and the host galaxy to explain our observations. Indeed, while there was indication of AGN contribution to the diagnostic diagram ratios in the East, the spaxels associated with the ring were consistent with the expected values for SF regions. Additionally, \cite{u_goals-jwst_2022} suggested that the outflow has a nearly face-on geometry, once again supporting the weak-coupling hypothesis. In fact, the presence of the powerful ring of star formation only a few hundreds of parsecs from the AGN is indication itself of the lack of globalized efficient feedback. 

\begin{figure}[]
\centering
\includegraphics[width=0.45\textwidth]{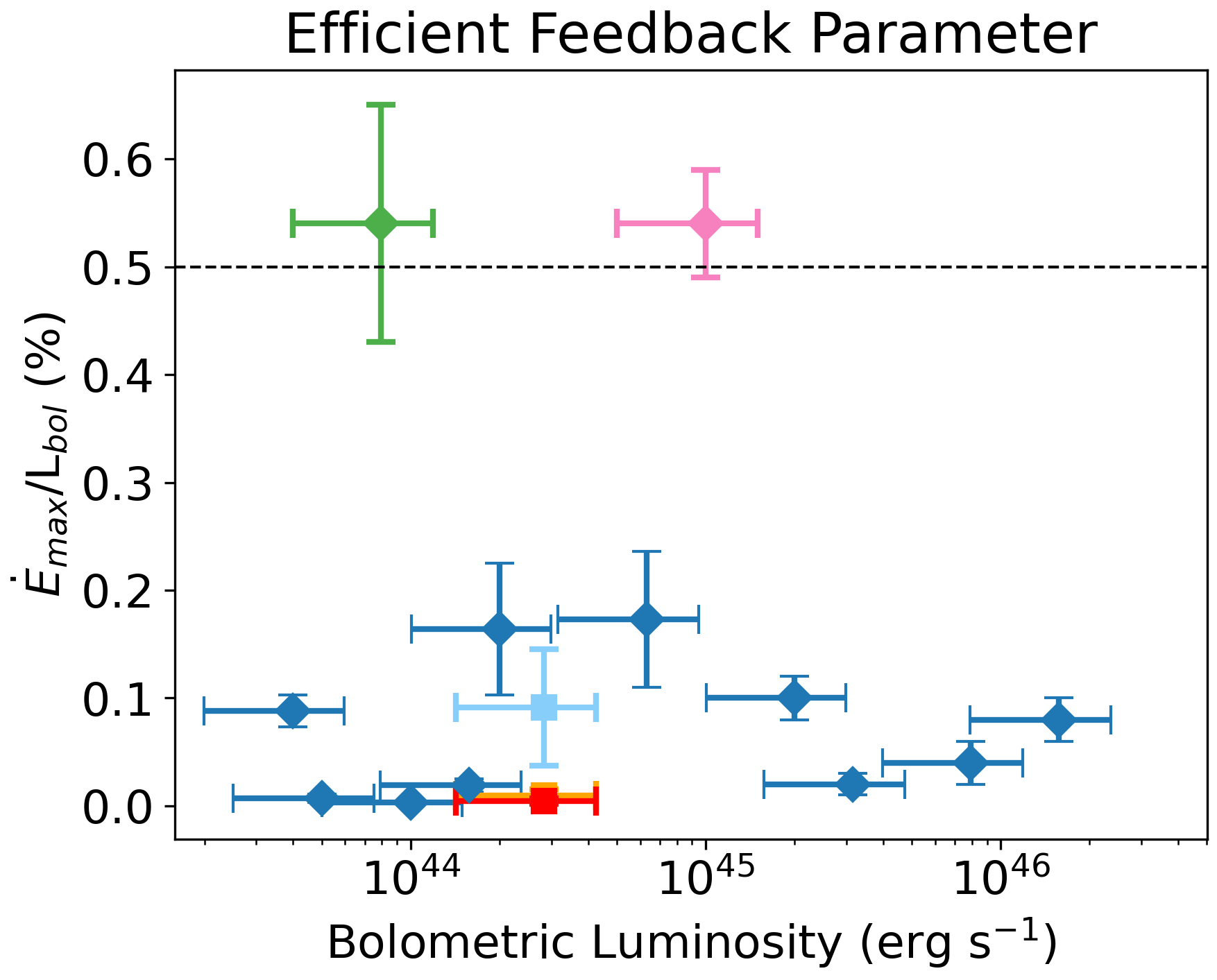}
    \caption{The efficient feedback parameter corresponds to the ratio of the maximum energy flow rate to the bolometric luminosity. According to \cite{Hopkins2010}, a ratio greater than 0.5\% is indicative of efficient feedback, which it indicated on the plot as a horizontal line. Symbols and colors as in Figure \ref{fig: outflow_params_compare}. 
    \label{fig: feedback_param}}
\end{figure}

The efficient feedback parameter is defined as the ratio of the energy flow rate and the bolometric luminosity of an AGN. This value is commonly given as a percentage, with a threshold of 0.5\% for a source to be considered to have efficient feedback \citep{Hopkins2010}. We plot this parameter for the sources outlined above as well as for our target in Figure \ref{fig: feedback_param}. Out of the 12 sources from \cite{Revalski2025} only two have values above the theoretical threshold: NGC~4151\footnote{Note that NGC 4151 is a highly variable source, and the efficient feedback parameter could be high because of a low value used for the bolometric luminosity and the issue of using non-contemporary measurements of L$_{bol}$ and $\dot{E}$.} and NGC~1068. It is important to note that this feedback parameter is not always a reliable measure of efficient feedback as it is based on calculations assuming that the black hole is radiating at the Eddington limit.

The efficient feedback parameters calculated for all three components in NGC~7469 all fall well below the 0.5\% mark but are consistent with other sources in the sample. Our value for the \texttt{nebular} component is closest to that of IC~3639. The two AGN with the lowest efficiency that were included in \cite{Fischer2013}, IC~3639 and NGC~788, were both classified as having ambiguous kinematics. NGC~7469 was classified as a compact source, similarly to NGC~5548, which is known for having its bicone pointed almost directly out of the plane of the host galaxy \citep{Crenshaw2009}. This suggests once again that the feedback processes in NGC~7469 are not strong enough to affect the surrounding medium in a significant way. 

While our measurements are not affected by the same issues as other studies which use global values, there are some assumptions that would affect our results for the efficient feedback parameter. For instance, using an AGN outflow inclination of 57$^{\circ}$ would increase the kinetic energy flow rate by 60\% \citep{MullerSanchez2011}. However, to get NGC~7469 close to or above the threshold for efficient feedback, we would need to increase the kinetic energy outflow rate by a factor of five. As the kinetic energy flow rate is proportional to the mass participating in the outflow and the cube of the velocity of the gas ($\dot{E} \propto M_{out}v_{out}^3$), this would require five times as much mass as is currently measured or increasing the velocities by a factor of 2. Increasing the opening angle or inclination of the bicone to increase the degree of coupling with the gas in the host would greatly affect the mass in the outflow, but the central engine would need to be much more powerful than it currently is to drive efficient feedback. While NGC~7469 radiates at 12.5\% Eddington \citep{Feuillet2025}, the bolometric luminosity would need to increase by a factor of 4 to sufficiently increase the outflow velocities ($v_{out} \propto \sqrt{L_{bol}}$, \citealp{Ramirez2012}). 


Our understanding of the origin of the X-ray gas is limited, especially at the sub-parsec scale, due to the resolution constraints of the central spaxel. Specifically, the innermost few parsecs of this galaxy might not contain enough gas to produce a powerful X-ray wind. However, upcoming XRISM observations (PI: G. Lanzuisi) may be able to reveal if there are any faster components associated with an inner accretion disk wind in NGC~7469. Additionally, there may not be a significant amount of material available to be entrained between the AGN and the surrounding ring, which could further impact the observed dynamics. The efficiency of outflows is likely connected to the evolutionary stage of the host galaxy, as suggested by comparisons with other Seyfert galaxies. Future research should focus on analyzing AGN with stronger X-ray outflows, which may represent a different phase in their evolution.

\section{Conclusions}

In this paper, we demonstrate the usefulness of IR footprint lines in tracing X-ray outflows in NGC~7469. We present a summary of our results below.

\begin{enumerate}
    \item We first obtained emission line fluxes in NGC~7469 using both X-ray and IR spectroscopic data. In order to identify which high-IP IR lines can trace the soft X-ray emitting gas, we employed \texttt{Cloudy} photoionization models and demonstrated the accuracy of these models to reproduce our observational results. We define three components corresponding to different phases of the outflow: \texttt{nebular}, \texttt{coronal}, and \texttt{footprint}.
    \item Using available Chandra ACIS narrow-band images centered around the three X-ray emission lines which have corresponding IR lines, we created contours to highlight the morphology of the emission and overplotted it onto the IR flux maps of our footprint lines. Despite the limitations of this comparison, we found general agreement between the morphology of the X-ray gas and the footprint line emission.
    \item We measured global kinematic information for the footprint and compared them to previous kinematic measurements of X-ray lines in NGC~7469. We found that the \osev velocity shift is very similar to that of the two corresponding footprint lines, suggesting a common origin.
    \item We used the spatial resolution of the JWST instruments to obtain mass and outflow parameter values for regions defined around the central AGN, using the measured flux and maximum velocity for the representative lines of each component. We find that the values are greatest to the East of the nucleus and for the X-ray gas component.
    \item We put the outflow parameters obtained for NGC~7469 in this paper into context by comparing our values to other low redshift Seyfert galaxies and to previous investigations of NGC~7469. We find that our values are in agreement with the previously observed trends. 
    \item The feedback parameters obtained for NGC~7469 are of 0.005\%, 0.01\% and 0.1\% for the \texttt{nebular}, \texttt{coronal}, and \texttt{footprint} components respectively. This places our AGN well below the threshold for efficient feedback but determines that the main dynamic drivers lies within the X-ray gas, and supports previous findings which suggested weak coupling between the AGN and the host galaxy. 
\end{enumerate}

Given the current limitations in spatial resolution achievable with Chandra, particularly in the soft X-ray range, it is important to emphasize the potential of using footprint lines to recover this spatial information. Furthermore, we have demonstrated that these footprint lines can also provide spatially resolved kinematic information when analyzed with JWST’s IFUs.

\begin{acknowledgments}
\added{The authors would like to thank the anonymous referee for their valuable comments, which have greatly improved this paper}. This work was partially supported by Chandra/SAO grant number GO3-24085X. This work is based on observations made with the NASA/ESA/CSA James Webb Space Telescope. The data were obtained from the Mikulski Archive for Space Telescopes at the Space Telescope Science Institute, which is operated by the Association of Universities for Research in Astronomy, Inc., under NASA contract NAS 5-03127 for JWST. This work is also based on observations obtained with XMM-Newton, an ESA science mission with instruments and contributions directly funded by ESA Member States and NASA. The specific observations analyzed can be accessed via MAST DOI: \dataset[10.17909/z7pe-qz75]{10.17909/z7pe-qz75} \citep{MAST_DOI}.
\end{acknowledgments}





%
\facilities{JWST(MIRI), JWST(NIRSpec), XMM-Newton(RGS), Chandra(HETG)}

\software{PySpecKit \citep{Pyspeckit2011, Pyspeckit2022},  
          Cloudy (v23.01; \citealp{Cloudy2023}), FTOOLS (v6.35.2; \citealp{Arnaud1996})
          XSPEC (v12.15.0; \citealp{Arnaud1996}), CIAO (v4.17;  \citealp{Fruscione2006}, QFitsView \citep{Ott2012}
          }


\appendix




\section{A Redshifted Outflow?}\label{app: redshifted outflow}

Outflowing winds in AGN have consistently been shown to be biconical in nature \citep{Pogge1988_bicone, Schmitt1994_bicone, Fischer2013}. Therefore, even if they are not detected due to observational limitations, a redshifted counterpart to the blueshifted outflow along our line of sight is expected. For this target, no concrete evidence of a redshifted outflow has previously been found, possibly because the redshifted part of the cone lies below the galaxy disk, where intervening dust from the host obscures our view. A tentative detection was reported by \citet{u_goals-jwst_2022} in [Mg~\textsc{v}] $\lambda$5.61\,\mic (IP~=~109\,eV) to the West of the nucleus, but the S/N was too low to draw any conclusions.

\begin{figure*}[h!]
\centering
\includegraphics[width=0.95\textwidth]{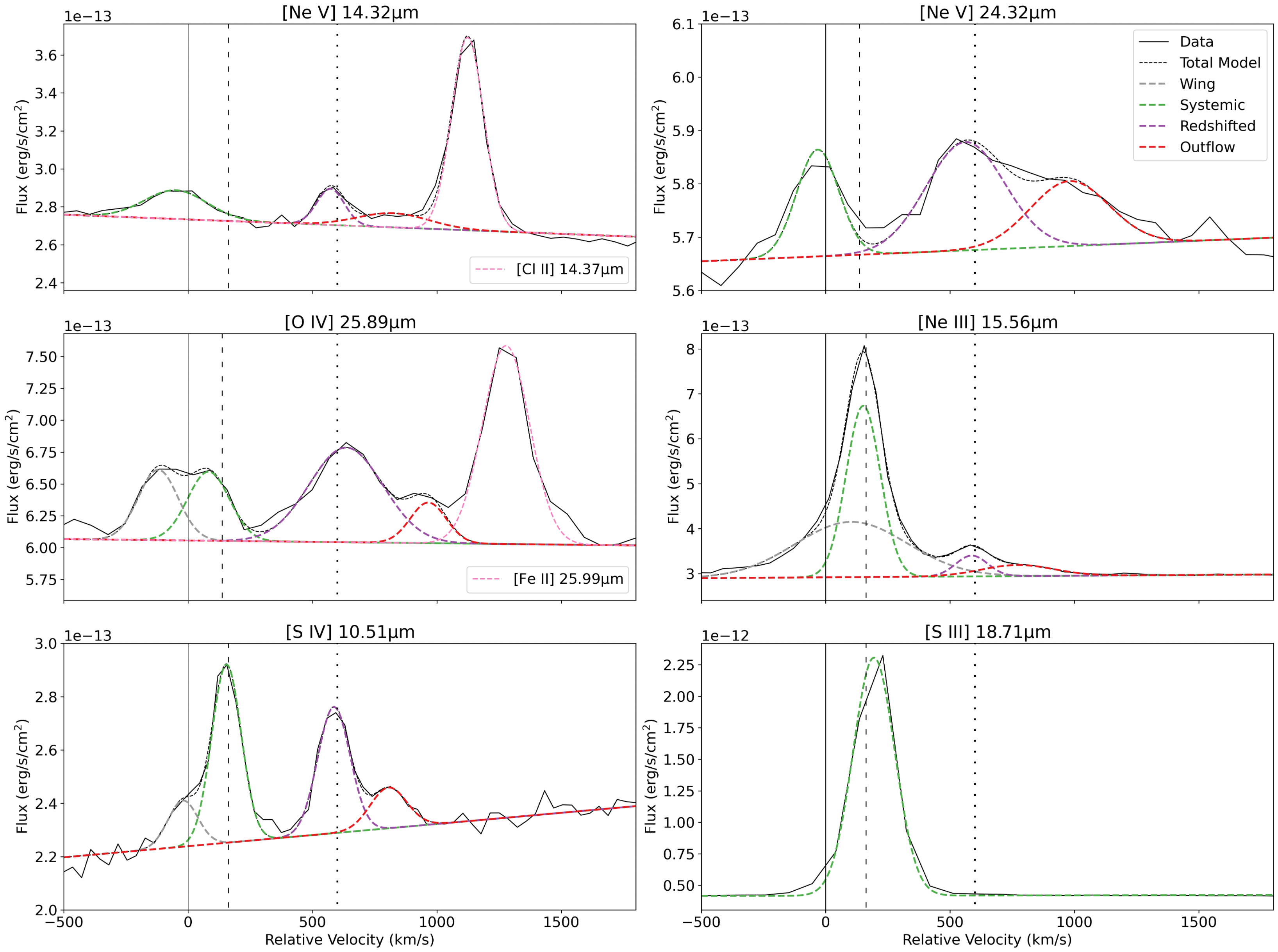}
    \caption{Spectra of the emission lines showing a redshifted component. We fit each spectrum with up to 5 Gaussians to account for asymmetry in both the systemic component and the redshifted component. The fifth component is used to fit the neighboring  [Cl~\textsc{ii}] $\lambda$14.37 \mic and [Fe~\textsc{ii}] $\lambda$25.99 \mic lines in the \nev $\lambda$14.32 \mic and \of $\lambda$25.89 \mic spectra respectively. The solid black vertical line highlights the systemic velocity, the dashed vertical line shows the rotational velocity values from \cite{cazzoli_ngc_2020}, and the dotted line is at a velocity of 600~km~s$^{-1}$ for easy visual comparison between panels. The outflowing gas is evident in the higher IP lines, but is absent from the star formation tracer \st $\lambda$18.71 \micns.  
    \label{fig: redshifted_outflow}}
\end{figure*}

We isolated a region in the northwest of the map (white dashed, Figure \ref{fig: regions}) and found strong indications of a redshifted outflow, as demonstrated in Figure~\ref{fig: redshifted_outflow}. This redshifted component is visible in high-ionization lines with sufficient S/N, as well as in lines with IPs as low as 34.8~eV, but it is completely absent in low-ionization star formation tracers such as \st (Figure~\ref{fig: redshifted_outflow}, bottom right panel). This redshifted component is detected in regions on the West side that are closer to the nucleus, however, in these regions the systemic component is strong enough that the outflow is barely noticeable in comparison. 

We fit up to 5 Gaussians to the emission lines and the redshifted component in our northwest region. It is difficult to draw any conclusions regarding a red wing in \nev due to the presence of the [Cl \textsc{ii}] $\lambda$14.37 \mic line, and a similar issue arises for \of $\lambda$25.89 \mic due to the neighboring [Fe~\textsc{ii}] $\lambda$25.99 \mic line. A strong red outflowing component to the redshifted emission is present in \nev $\lambda$24.32 \mic but the long wavelength means that the line is subjected to strong fringing issues. However, the outflow is clearly present in both \sfo $\lambda$10.52 \mic and \net $\lambda$15.56 \micns. 

We overplotted vertical lines corresponding to the rotational velocities of the thin ionized gas disk (163~km~s$^{-1}$) and the dynamically hotter component (137~km~s$^{-1}$) reported by \citet{cazzoli_ngc_2020}, and find good agreement with our observations. The velocity of the redshifted component ranges between 561 and 634~km~s$^{-1}$, while the red outflow shift ranges from 772 to 980~km~s$^{-1}$. This velocity shift is generally consistent with the velocities found in the blueshifted side of the bicone in the X-ray (-660~km~s$^{-1}$, \citealp{mehdipour_multi-wavelength_2018, grafton-waters_multi-wavelength_2020}) and in the optical (-715~km~s$^{-1}$, \citealp{robleto-orus_muse_2021, xu_spatially_2022}), further supporting the hypothesis that this component is related to the AGN outflow. 

We measured the fluxes of both the total emission and the redshifted component alone. Using a \texttt{Cloudy} model with log(U)~=~-1.5, we confirmed that the AGN can ionize gas at this distance.\footnote{We also examined the corresponding region in the NIRSpec data for mid-ionization lines, [Mg~\textsc{iv}] $\lambda$4.49 \mic (IP~=~80.1~eV) and [Ar~\textsc{vi}] $\lambda$4.53 \mic (IP~=~74.8~eV). However, the S/N at this location was too low to draw firm conclusions. This is consistent with our \texttt{Cloudy} models, which predicted fluxes for these lines to be an order of magnitude lower than the faintest line detected with MIRI.} We then used the measured fluxes to place this region on the mid-IR diagnostic diagrams of \citet{Feuillet2025} to assess the nature of the emission (see Table \ref{tab: redshifted component}). When considering the total flux, the line ratios place the region firmly within the star-forming (SF) regime. This is expected, given the region’s proximity to the ring and the tendency for weak AGN signatures to be overwhelmed by star formation in diagnostic diagrams \citep{Melendez2008-AGN-Contribution, feuillet_classifying_2024}. However, when isolating only the redshifted component of each emission line, the resulting ratios are consistent with AGN activity. Notably, we could not compute line ratios that include star formation tracers for the redshifted component, as these lines showed no detectable redshifted emission. This absence further supports an AGN origin for the redshifted emission.

\begin{table}[h!]
\begin{center}
\caption{Diagnostic ratio values for the total and redshifted components of the emission lines.}
\label{tab: redshifted component}
\begin{tabular}{ccc}
\hline
Line Ratio & Total & Redshifted \\
\hline\hline
log \netns/\netw      & -0.80 $\rightarrow$ SF & - \\
log \sfons/\st        & -1.48 $\rightarrow$ SF & - \\
log \sfons/\net       & -0.99 $\rightarrow$ SF & -0.37 $\rightarrow$ AGN \\
log \nevns14/\net     & -1.21 $\rightarrow$ SF & -0.52 $\rightarrow$ AGN \\
\hline
\end{tabular}
\end{center}
\end{table}



\section{Cloudy Model Parameters}\label{app: mod_params}

\begin{table}[h!]
\centering
\caption{Cloudy Model Parameters used in the mass calculations.}
\label{tab: Cloudy model params}
\begin{tabular}{ccccccc}
\hline
r & Parameter & log(U) &  n$_H$ & N$_H$ & $\delta$    & A$_{max}$  \\
pc &          &        & log cm$^{-3}$ & log cm$^{-2}$ & log cm & log cm$^{-2}$ \\
(1)  &           &       & (2)       & (3) & (4) & (5)  \\
\hline\hline
159 & Footprint & 0     & 1.14   & 21.83  & 20.69 & 41.40    \\
159 & Coronal   & -1.25 & 2.39   & 21.16  & 18.77 & 41.40    \\
159 & Nebular   & -2.25 & 3.39   & 20.16  & 16.77 & 41.40    \\
318 & Footprint & 0     & 0.53   & 21.22  & 20.69 & 42.00    \\
318 & Coronal   & -1.25 & 1.78   & 20.55  & 18.77 &  42.00    \\
318 & Nebular   & -2.25 & 2.78   & 19.55  & 16.77 & 42.00         \\
424 & Footprint & 0     & 0.28   & 20.97  & 20.69 & 42.25    \\
424 & Coronal   & -1.25 & 1.53   & 20.30  & 18.77 &  42.25        \\
424 & Nebular   & -2.25 & 2.53   & 19.30  & 16.77 &  42.25        \\
583 & Footprint & 0     & 0.01   & 20.70  & 20.69 & 42.53    \\
583 & Coronal   & -1.25 & 1.26   & 20.03  & 18.77 &  42.53        \\
583 & Nebular   & -2.25 & 2.26   & 19.03  & 16.77 &  42.53        \\
742 & Footprint & 0     & -0.20  & 20.49  & 20.69 & 42.74    \\
742 & Coronal   & -1.25 & 1.05   & 19.82  & 18.77 & 42.74         \\
742 & Nebular   & -2.25 & 2.05   & 18.82  & 16.77 &  42.74        \\
848 & Footprint & 0     & -0.32  & 20.37  & 20.69 & 42.86    \\
848 & Coronal   & -1.25 & 0.93   & 19.70  & 18.77 & 42.86         \\
848 & Nebular   & -2.25 & 1.93   & 18.70  & 16.77 &  42.86       \\    
\hline
\end{tabular}
\tablecomments{(1) r is the distance in pc from the center of the central spaxel to the center of the region. (2) n$_H$ is the density of the gas calculated from U and r using Equation \ref{eq: U}. (3) N$_H$ is the column density set such that it does not cause the model thickness to exceed the maximum thickness. (4) $\delta$ is the model thickness calculated from the density and the column density. For reference, the maximum thickness for all regions is of log $\delta_{max}$\,=\,20.69\,cm. (6) A$_{max}$ is the maximum emitting area allowed by the size of the region.}
\end{table}


\bibliography{sample7}{}
\bibliographystyle{aasjournalv7}



\end{document}